\documentclass[fleqn,usenatbib]{mnras}

\usepackage[T1]{fontenc}

\DeclareRobustCommand{\VAN}[3]{#2}
\let\VANthebibliography\thebibliography
\def\thebibliography{\DeclareRobustCommand{\VAN}[3]{##3}\VANthebibliography}

\usepackage{graphicx}	
\usepackage{newtxtext,newtxmath}

\newcommand{\pderiv}[2]{\frac{\partial #1}{\partial #2}}
\newcommand{\rc}{\color{black}}

\title[Discs in the Galactic Centre]{The evolution of circumstellar discs in the Galactic Centre: an application to the G-clouds}

\author[Owen, J. E. \& Lin, D. N. C.]{
James E. Owen$^{1}$\thanks{E-mail: james.owen@imperial.ac.uk} and Douglas N. C. Lin $^{2, 3}$\\
$^{1}$Astrophysics Group, Imperial College London, Blackett Laboratory, Prince Consort Road, London SW7 2AZ, UK\\
$^{2}$ Department of Astronomy and Astrophysics, University of California, Santa Cruz, CA 95064, USA\\
$^{3}$Institute for Advanced Studies, Tsinghua University, Beijing, 100086, China
}

\date{Accepted XXX. Received YYY; in original form ZZZ}

\pubyear{2015}

\begin{document}
\label{firstpage}
\pagerange{\pageref{firstpage}--\pageref{lastpage}}
\maketitle

\begin{abstract}

The Galactic Centre is known to have undergone a recent star formation episode a few Myrs ago,  which likely produced many T Tauri stars hosting circumstellar discs. It has been suggested that these discs may be the compact and dusty ionized sources identified as ``G-clouds''. Given the Galactic Centre's hostile environment, we study the possible evolutionary pathways these discs experience. We compute new external photoevaporation models applicable to discs in the Galactic Centre that account for the sub-sonic launching of the wind and absorption of UV photons by dust. Using evolutionary disc calculations, we find that photoevaporation's rapid truncation of the disc causes them to accrete onto the central star rapidly. Ultimately, an accreting circumstellar disc has a lifetime $\lesssim1~$Myr, which would fail to live long enough to explain the G-clouds. However, we identify a new evolutionary pathway for circumstellar discs in the Galactic Centre. Removal of disc material by photoevaporation prevents the young star from spinning down due to magnetic braking, ultimately causing the rapidly spinning young star to torque the disc into a ``decretion disc'' state which prevents accretion. At the same time, any planetary companion in the disc will trap dust outside its orbit, shutting down photoevaporation. The disc can survive for up to $\sim$10 Myr in this state. Encounters with other stars are likely to remove the planet on Myr timescales, causing photoevaporation to restart, giving rise to a G-cloud signature. A giant planet fraction of $\sim10\%$ can explain the number of observed G-clouds.

\end{abstract}

\begin{keywords}
Galaxy: centre -- stars: pre-main-sequence -- protoplanetary discs -- accretion, accretion discs -- black hole physics
\end{keywords}



\section{Introduction}
The centre of our galaxy, the Milky Way, contains a supermassive black hole (Sgr A*). Through the analysis of the orbits of stars, either statistically or individually, in the vicinity of Sgr A*, its mass has been determined to be $\sim 4\times 10^{6}$~M$_\odot$ \citep[e.g.][]{Eckart1997,Ghez1998,Ghez2003,Gillessen2009,Genzel2010,Gravity2018}. As such, the Galactic Centre provides a unique window into astrophysical processes that are prevalent in the majority of galaxies. The Galactic Centre contains a nuclear star cluster \citep[e.g.][]{Genzel1987,Rieke1988,Ghez2005,Genzel2010}, which contains a large number of massive and hot stars indicative of recent star formation \citep[e.g.][]{Krabbe1995,Paumard2001,Ghez2003,Eisenhauer2005,Paumard2006,Bartko2009,Do2013}. Follow-up analysis suggests that the most recent star formation event occurred 3-5 Myr ago, giving birth to a large number of massive stars with a top-heavy IMF \citep[e.g.][]{Paumard2006,Bartko2010,Lu2013}.

Detailed analysis of the Galactic Centre with near diffraction-limited observations \citep[e.g.][]{Ghez2005b,Gillessen2009} has produced some unexpected results. \citet{Gillessen2012} reported the discovery of a compact, dusty ionized cloud, G2. Multi-epoch observations allowed its orbit around Sgr A* to be determined, showing this object was on an extremely eccentric orbit \citep[e.g.][]{Gillessen2012,Gillessen2013,Pfuhl2015,Plewa2017} and connections were made to an earlier detected cloud, G1, \citep[e.g.][]{Clenet2005}. Recently, additional objects have been discovered \citep{Eckart2013,Ciurlo2020,Peissker2020} taking the number of known ``G-clouds\footnote{Throughout the literature, these objects have been given different nomenclature (G-clouds, DSOs, D-clouds); in this work, we simply refer to them as G-clouds.}'' to around 5-10. The more recently detected objects do not possess the same extreme eccentricity as G1/G2 and do not appear to be located in a common orbital plane \citep{Ciurlo2020}. The objects show similar observational characteristics: NIR emission with a black-body temperature of $\sim 600$~K and luminosities less than a few times Solar, indicative of dust, and Br-$\gamma$/HeI emission indicating the presence of ionized gas with a temperature of $\sim 10^{4}$~K, suggesting this gas is photoionized by the EUV background in the Galactic Centre \citep[e.g.][]{Gillessen2012}. This recombination emission implies densities of $\sim 10^5-10^6$~cm$^{-3}$ on size scales of 10-100 AU \citep[e.g.][]{Gillessen2012}. 

Despite a decade of study, the origin and astrophysical nature of the G-clouds remain a mystery. \citet{Burkert2012} discusses two possible classes: a diffuse cloud scenario, where the G-clouds are pressure-confined clouds, or a compact source scenario, where the G-clouds represent the diffuse gas and dust surrounding an undetected central mass. Simulations of the pressure-confined diffuse gas cloud scenario showed agreement with the position-velocity properties of the Br-$\gamma$ emission \citep[e.g.][]{Burkert2012,Schartmann2012,Schartmann2015}. However, these simulations demonstrated that, due to the strong tidal forces from Sgr A*, these diffuse gas clouds could not survive pericentre passage if they were on an extremely eccentric orbit like G2. During its pericentre passage in 2014, observations of G2 \citep{Plewa2017} showed that it did survive pericentre passage, although becoming tidally distorted on its closet approach. The fact that G2 (and G1) did survive pericentre passage strengthens the case for the compact-source scenario. 

Several different versions of the compact source scenario have been presented: photoevaporating circumstellar discs, either around T Tauri stars  \citep{Murray-Clay2012}, or created through tidal interactions \citep{MiraldaEscude2012}; a nova \citep{Meyer2012}; winds launched from young T Tauri stars \citep[e.g.][]{Scoville2013,Ballone2013,Ballone2018,DeColle2014,Valencia2015,Zajacek2017} or merger products \citep[e.g.][]{Witzel2014,Prodan2015}. 

There have been claims of a central source detected in the K-band for G2 \citep[e.g.][]{Peissker2020,Peissker2021} indicative of a young T Tauri star; however, these claims have been disputed, with other authors finding a non-detection \citep[e.g.][]{Gillessen2012,Pfuhl2015,Plewa2017}. Nevertheless, the constraints on the central luminosity of any T Tauri star are weak, with luminosities $\lesssim 5$~L$_\odot$ consistent with the observations \citep[e.g.][]{Gillessen2012}. A photoevaporating circumstellar disc is appealing as the origin of the G-clouds. The Galactic Centre is known to be a site of star formation \citep[e.g.][]{Genzel2010}, which naturally includes the emergence of T Tauri stars.  The recent star formation episode could have produced, at maximum, a few thousand T Tauri stars \citep[e.g.][]{Nayakshin2005,Lu2013}, which would initially host circumstellar discs. The background UV environment of the Galactic Centre produces photoevaporation rates and densities consistent with the observational constraints. This agreement led \citet{Murray-Clay2012} to propose that G2 is a protoplanetary disc of size $\sim 10 $AU that is being photoevaporated and tidally stripped as it approached Sgr A*.  With an alternative scenario, \citet{MiraldaEscude2012} proposed that G2 was a photoevaporating circumstellar disc created by a close encounter between a black hole and a star. While these early works were suggestive, they did not precisely address the evolution of such photoevaporating discs, nor did they use detailed photoevaporation models. In particular, previous works have ignored the ongoing accretion process, which channels gas from the discs to their central hosts. Typically, accretion plays an important (if not the dominant) role in the removal of disc material, not photoevaporation \citep[e.g.][]{Clarke2007}. Therefore, simply comparing photoevaporation rates to disc masses (as done in previous work) is likely to overestimate their lifetimes. Additionally, photoevaporation causes disc radii to shrink to the point where the photoionized gas is formally bound to the host star. It is well-known photoevaporation can still occur, driven by sub-sonic pressure gradients in the bound gas \citep[e.g.][]{Adams2004}. However, no photoevaporation model applicable in this bound limit has been developed for the high-EUV fluxes prevalent in the Galactic Centre. Thus, in this work, we study the properties and evolution of circumstellar discs in the Galactic Centre by (i), developing a photoevaporation model applicable to circumstellar discs in the Galactic Centre and (ii), considering the secular evolution of circumstellar discs through numerical computation. In doing this work, we also use our results to critically assess the photoevaporating disc scenario for the G-clouds. 

\section{The radiation environment}\label{sec:radiation}
The photoevaporation of objects is controlled by the amount of UV radiation they receive. In the Galactic Centre, as we shall demonstrate explicitly later, the UV radiation environment is so strong that ionizing EUV radiation plays the controlling role. For example, the EUV luminosity emanating from the central parsec is $\sim 10^{51}$ ph~s$^{-1}$ \citep[e.g.][]{Martins2007}; however, this arises from the combination of numerous sources and will vary with position. Therefore, in order to calculate the evolution of a photoevaporating object, we must estimate the EUV flux as a function of position in the galactic centre. 

There are two populations of young stars that are sufficiently massive to emit large amounts of ionizing radiation. First, the S-star cluster, which is dominated by main sequence B-stars \citep[e.g.][]{Schodel2003,Ghez2003} and is approximately spherically symmetric \citep[e.g.][]{Genzel2010}. The second population is the O/WR stars, this stellar cluster is located in a ring \citep[e.g.][]{Levin2003} with an inner edge of $\sim 0.04~$pc and an outer edge of $\sim 0.5$~pc \citep[e.g.][]{Paumard2006,Lu2009,Bartko2009}. It is currently debated whether there are one or multiple rings, or the rings are warped \citep[e.g.][]{Genzel2003,Paumard2006,Lu2009,Cuadra2008,Bartko2010,Naoz2018}; however, in our calculation, we simply assume a single ring and calculate EUV fluxes within and perpendicular to the disc's plane. This approach then covers the range of expected EUV fluxes for multiple rings. 

Since the number of ionizing sources is not very large ($N_*$ is a few hundred), the EUV fluxes as a function of distance from Sgr A* can vary due to Poisson noise in the exact positions of all the individual stars. {\rc Additionally, the use of smooth density distributions is sensitive to any chosen smoothing length due to the small number of stars}.  Thus, we choose to estimate the EUV fluxes through Monte Carlo modelling of the distribution of the star's positions, masses and ionizing luminosities. Each Monte Carlo realisation consists of multiple steps. Firstly, based on observed stellar surface density distribution of $\Sigma_S\propto R^{-1.5}$  and $\Sigma_O\propto R^{-1}$ for the S-star cluster and O/WR disc respectively \citep[e.g.][]{Bartko2010,Genzel2010,Do2013} we randomly draw positions for individual stars. We assume the S-star cluster is spherically symmetric and consider stars out to distances from Sgr A* of 1~pc. For the O/WR disc, we assume it extends from 0.04 to 0.5~pc and has a constant thickness of $H/R=0.2$\footnote{Provided $H/R\ll1$ we find its choice does not affect our results.}. For the S-stars, we consider 110 stars (the value obtained by integrating the observed surface density profile out to 1~pc, and comparable to the 112 stars identified by \citealt{Ali2020}), and for the O/WR disc, we use the number required {\rc (typically $\sim$250 stars)} to give an initial stellar mass cluster of $10^{4}$~M$_\odot$ \citep{Lu2013},. 

The next step is to find the star's initial mass. For this, we draw from the measured IMFs for the two clusters. For the S-stars, we use a Salpeter IMF \citep[e.g.][]{Bartko2010}. The O/WR disc is thought to have a top-heavy mass function \citep[e.g.][]{Nayakshin2005,Paumard2006}, thus we use a top-heavy IMF of ${\rm d}N_*/{\rm d}M_*\propto M_*^{-1.7}$ \citep[with $M_*$ the stellar mass, e.g.][]{Lu2013}. With our random sampling, we exclude stars less massive than 8~M$_\odot$ as they are not observable (nor do they contribute to the EUV flux), and we exclude stars more massive than 150~M$_\odot$ as they were not included in the young star properties estimated by \citet{Lu2013}. Based on the initial masses, we use a $\sim 3$~Myr isochrone to extract the current stellar properties (approximately the age of the young stars from \citealt{Lu2013}). This 3~Myr isochrone is obtained from the {\sc mist} stellar tracks \citep{Dotter2016,Choi2016}, created using {\sc mesa} \citep{Paxton2011,Paxton2013,Paxton2015} for solar metallicity stars\footnote{The \citet{Lu2013} characterisation of the stellar population, which we follow, used solar-metallicity tracks.} with initial rotation rates of $v/v_{\rm crit}=0.4$ (the default value). We then use the stellar properties to estimate the star's EUV Luminosity using the stellar atmospheres of \citet{Thompson1984}, which were tabulated and displayed graphically in Figure~4 of \citet{Kuiper2020}, accounting for absorption of EUV photons in the stellar atmosphere.

Finally, we include attenuation of the EUV photons between stars due to dust in the Galactic Centre. Throughout this work, we adopt an EUV dust opacity\footnote{We specifically mean per unit mass of dust, rather than the standard quoted value of per mass of a gas and dust mixture with a specified dust-to-gas ratio. This is because our dust-to-gas mass ratio will vary later.\label{fot:dust-foot}} of $\kappa_{\rm d,EUV} = 10^{5}~$cm$^2$~g$^{-1}$, appropriate for ionizing photons being absorbed by small sub-micron sized silicate dust particles \citep{Ali2021}\footnote{Note \citet{Ali2021} present their dust opacity in terms of the dust and gas mixture, see footnote~\ref{fot:dust-foot}.}.  Since the nature of the dust-and-gas distribution in the galactic centre remains uncertain we adopt a dust-to-gas mass ratio of 0.01 and consider a gas distribution of $n(r)=n_0(r/r_0)^{-\gamma}$ where $n_0 = 10^{4}$~cm$^{-3}$ and $r_0 = 10^{15}$~cm. We take either $\gamma = -1$ as proposed by \citet{Xu2006} or adopt a flatter distribution $\gamma=-0.5$ as proposed by \citet{Gillessen2019}. 

\begin{figure}
    \centering
    \includegraphics[width=\columnwidth]{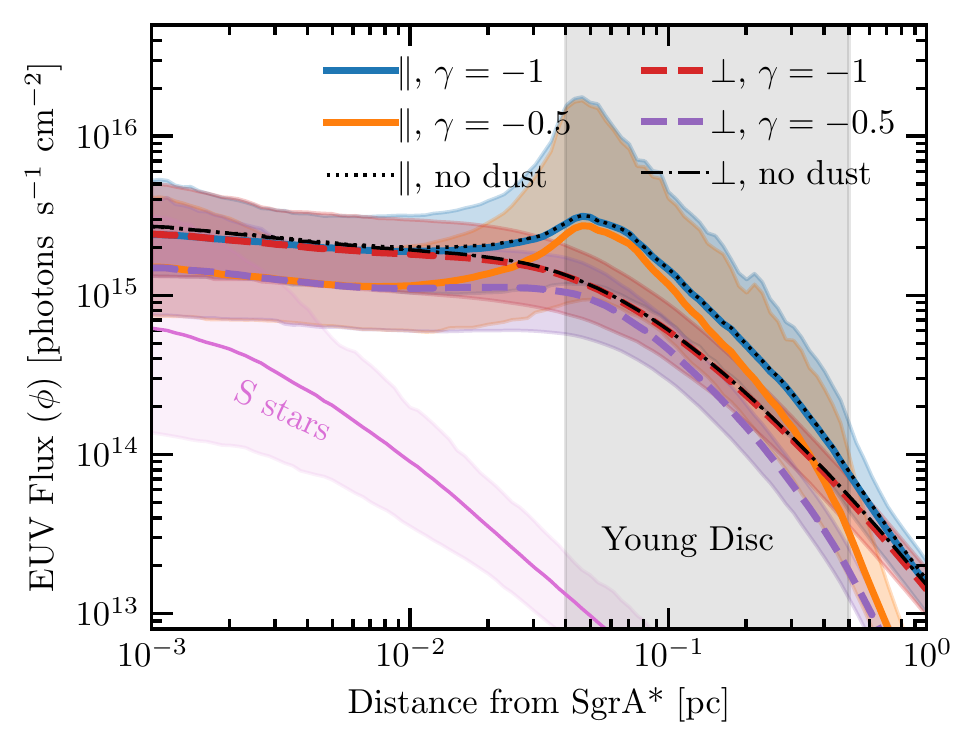}
    \caption{The lines show the median EUV flux as a function of distance from the Galactic Centre. The lines for the perpendicular models are for the EUV flux perpendicular to the plane of the O/B young discs and the parallel models are for the EUV flux in the plane of the disc. The different $\gamma$'s show the effect of attenuation due to dust in the Galactic centre for two profiles with $n\propto r^{-\gamma}$. The shaded region around the lines represents the 5--95\% range from the realisations. The region of the young disc is shown shaded in grey. }
    \label{fig:EUV_flux}
\end{figure}

Based on this dust distribution and the positions and luminosities of all the stars for a given realisation, we then use ray tracing to compute the EUV flux impinging on a specific point at a known distance from Sgr A*. This point is taken to either lie in the plane of the O/WR stellar disc or perpendicular to it. We repeat this analysis 500 times to extract the expected range of EUV fluxes as a function of distance from Sgr A*. The result of this exercise is shown in  Figure~\ref{fig:EUV_flux}. 

This analysis shows that the massive stars in the O/WR young disc dominate the EUV flux throughout the Galactic Centre. This results in an essentially flat EUV flux distribution inside the inner edge of the disc. 
These fluxes can be compared to the extreme end of the EUV fluxes experienced by the ``proplyds'' (known photoevaporating protoplanetary discs) in the Orion Nebula Cluster of $10^{13}$ to a few $10^{14}$~photons~s$^{-1}$~cm$^{-2}$ \citep{Bally1998,Johnstone1998}. Our model produces an EUV luminosity emanating from the inner parsec ranging from  $5\times10^{50}-3\times10^{51}$~photons~s$^{-1}$ in agreement with the observational estimate of $\sim 10^{51}$~photons~s$^{-1}$ \citep[e.g.][]{Martins2007,Genzel2010}. Our result shows that dust attenuation plays a small role, only reducing the fluxes by a factor of two in the inner regions. Finally, we do note that our Monte-Carlo sampling does indicate cases where an individual star is nearby and completely dominates the EUV flux, resulting in fluxes that can exceed $10^{17}$ ~photons~s$^{-1}$~cm$^{-2}$. If this were to happen, it was likely to be only short-lived. In the inner 0.1~pc, the average distance between stars is $\sim 10^{16}$~cm. Thus, the travel time to move out of the range of a single star and closer to the average is $t_{\rm travel}\sim 10^{16} {\rm ~cm}/(4\times10^{7}~{\rm cm~ s^{-1}})\sim 10$~years, short compared to the lifetime of any star/object in the galactic centre. In general, we expect the EUV flux received at a given point in the Galactic Centre to be dominated by the combination of a large fraction of the massive stars in the O/WR disc. We will take our nominal flux to be $3\times10^{15}$~photons~s$^{-1}$~cm$^{-2}$ and consider a low value of $10^{14}$~photons~s$^{-1}$~cm$^{-2}$.

\section{The anatomy of the evaporating disc}
Our goal is to explore the photoevaporating disc scenario as the origin of the G-clouds in detail. To achieve this goal, we need to know the photoevaporative mass-loss rates as a function of star and disc parameters, as well as how they vary with the UV environment.  However, the EUV fluxes we have derived for the Galactic Centre in the previous section exceed those of typical star-forming regions. As discussed in the Introduction, such modern external photoevaporation calculations are not applicable to this extreme environment. Thus, we calculate a new family of external photoevaporation models in this section. {\rc Additionally, to launch a photoevaporative outflow, the external pressure needs to be sufficiently low to allow the wind to be accelerated to super-sonic speeds. While this is generally true in the Galactic Centre, as we discuss in Section~\ref{sec:pressure} it is not true near Sgr A* (and hence for G2 near pericentre). }

We calculate the photoevaporation models using a modified version of the \citet{Johnstone1998} framework. While there has been significant recent work advancing external photoevaporation calculations \citep[e.g.][]{Adams2004,Facchini2016, Haworth2018,Haworth2019,Winter2022} the focus has been to improve the calculations at lower UV irradiation levels, more appropriate for typical galactic star-forming environments \citep[e.g.][]{Fatuzzo2008,Adams2010}. In these cases, the EUV irradiation is unable to penetrate deep into the FUV heated region and does not affect the dynamics \citep[e.g.][]{Adams2004,Haworth2018}. {\rc However, at high EUV fluxes, the ionization front is able to penetrate below where the sonic point in a pure FUV heated flow would be. Thus,}
the flow structure is sub-sonic in the FUV heated region, and the EUV flux controls the outflow properties \citep[e.g.][]{Johnstone1998}. {\rc In the Appendix, we show that the EUV fluxes in the Galactic Centre are significantly above the critical flux required for the outflow to be in the EUV-controlled limit.} 

The original \citet{Johnstone1998} formalism did not include thermal launching of the wind inside the ``critical-radius'', $R_g\equiv GM_*/c_s^2$, where $c_s$ is the isothermal sound speed \citep[e.g.][]{Adams2004}. This radius is approximately $10$~AU for a Solar mass star when the EUV heated region reaches $10^4$~K. As described above, it is possible the G-clouds can be smaller than this value, and as we show below, discs are likely to be rapidly evaporated to sizes smaller than this radius.  While EUV heated gas in the outer disc is still formally bound inside $R_g$, provided that there's sufficient energy input (from an external radiation field), thermally driven outflows can be launched from inside $R_g$ \citep{Parker1965,Lamers1999}. Previous work on disc evaporation in the Galactic Centre \citep[e.g.][]{Murray-Clay2012,MiraldaEscude2012} applied the \citet{Johnstone1998} model to discs with radii below $R_g$, even though this approach does not give the correct flow structure. Nor did these approaches include the FUV-heated sub-sonic layer.

Therefore, in this section, we update the theory of externally photoevaporating discs in the EUV regime to include wind launching inside $R_g$. In addition, we account for the attenuation of the UV radiation field due to dust. At dust-to-gas ratios of order $\sim 0.01$, absorption of EUV photons by neutral hydrogen typically dominates over dust in ionized gas. However, it can become important at larger dust-to-gas ratios. In some of our evolutionary simulations, we find the outer regions of our discs can reach dust-to-gas ratios approaching unity, and therefore we cannot neglect the absorption of EUV photons by dust. These complications increase the complexity of the models; however, to quickly scan the parameter space, we still adopt the \citet{Johnstone1998} formalism as the basis of our calculation. 

The flow topology is taken as follows: from the disc, an isothermal FUV heated outflow is launched that remains sub-sonic, the flow then transitions through a thin ionization front and becomes an ionized isothermal outflow. We take the sound speed of the gas to be 3 km~s$^{-1}$ in the FUV heated outflow, appropriate for gas in strong FUV fields \citep[e.g.][]{Hollenbach1999,Adams2004} and 10 km~s$^{-1}$ in the EUV regions, appropriate for ionized gas at 10$^4$~K. We assume the EUV outflow has spherical velocity divergence and the velocity and density profile follow an isothermal Parker wind \citep[e.g.][]{Parker1958,Cranmer2004}. We modify the isothermal sound speed of the outflowing fluid (of gas and dust) to be $c_{s,{\rm fluid}}=c_{s,{\rm gas}}/\sqrt{1+X_d}$ where $X_d$ is the dust-to-gas mass ratio in the wind, this accounts for the fact the effective pressure felt by a dusty-fluid is lower than one without dust. As it is the small dust grains that dominate the EUV opacity, we assume that the dust particles are 0.1$\mu$m. Dust grains of this size are easily entrained in photoevaporative outflows \citep{Facchini2016,Owen2021,Winter2022}, hence we take the dust-to-gas mass ratio to be constant throughout the outflow.  

\begin{figure}
    \centering
    \includegraphics[width=\columnwidth]{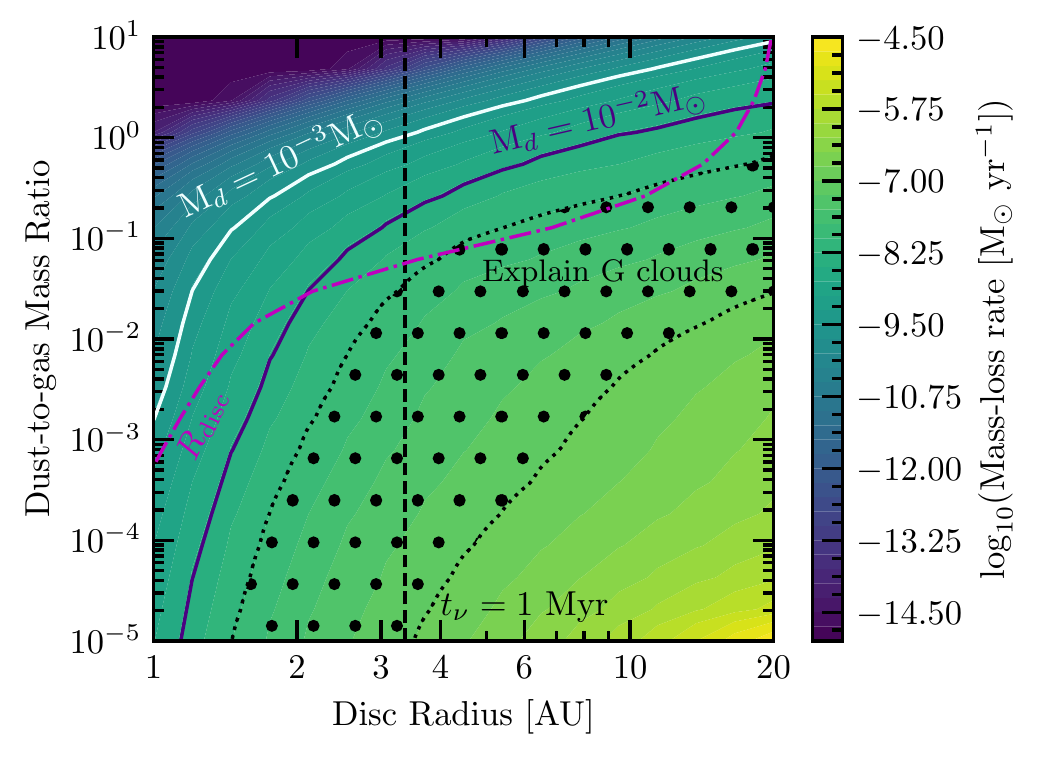}
    \caption{The external photoevaporation rates as a function of disc size and dust-to-gas mass ratio at the outer edge of the disc for an EUV flux of $3\times10^{15}$ photons s$^{-1}$. The photoevaporating mass-loss rates required to approximately explain the G-clouds are shown as the open-dotted region, with mass-loss rates between $10^{-8}$ and $10^{-7}$~M$_\odot$~yr$^{-1}$. Curves of constant disc survival time (3~Myr) are shown for disc mass of of $10^{-3}$ and $10^{-2}$~M$_\odot$ and a viscous time of 1~Myr (for $\alpha = 10^{-3}$) are shown. The dot-dashed line (labeled $R_{\rm disc}$) shows the radius an initially 0.1~M$_\odot$ disc would be truncated to in 3~Myr by photoevaporation alone.   As discussed in Section~\ref{sec:standard}, this figure demonstrates the inherent problem of having a standard protoplanetary disc explain the G clouds. A large disc requires an extremely high dust-to-gas ratio to survive photoevaporation. A small disc (to the left of the $t_\nu = 1$~Myr) line will drain onto the star via accretion rapidly.  Hence, for nominal dust-to-gas ratios, it is impossible to satisfy both constraints simultaneously, let alone explain the required photoevaporation rates for the G-Clouds.}
    \label{fig:photo_rates}
\end{figure}

The FUV heated region is strongly sub-sonic ($R_g$ for FUV heated gas is $\sim 10$ times larger than for EUV heated gas), hence we model the density structure in this region as being hydrostatic, including rotational support at constant specific angular momentum. Thus, the density profile in the FUV heated region, as a function of distance ($r$) becomes:
\begin{equation}
    \rho_{\rm FUV} = \rho_{d}\exp\left[\frac{GM_*}{c_{s,FUV}^2}\left(\frac{1}{r} - \frac{1}{R_d}\right) -\frac{h_d^2}{2 c_{s,FUV}^2}\left(\frac{1}{r^2}-\frac{1}{R_d^2}\right)\right] \label{eqn:hydrostatic}
\end{equation}
where $\rho_d$ is the density in the FUV heated region at the disc's outer edge and $h_d$ is the specific angular momentum in the outflow, given by $h_d=\sqrt{GM_*R_d}$. The fluxes for the EUV heated region are sufficiently high that the outflow is in ionization-recombination balance. As we include dust attenuation, we cannot simply make a Str\"omgren volume argument like \citet{Bertoldi1990,Johnstone1998}, and we must calculate the ionization structure explicitly. Making the on-the-spot approximation {\rc for pure hydrogen} we calculate the ionization fraction ($X_i$) as:
\begin{equation}
    X_i = \frac{1}{2}\left[\sqrt{\left(\frac{\Gamma}{\alpha_b n_H}\right)^2+4\frac{\Gamma}{\alpha_b n_H}} -\frac{\Gamma}{\alpha_b n_H} \right]
\end{equation}
where $n_H$ is the density of hydrogen, $\alpha_b$ is the case-B recombination coefficient, and $\Gamma$ is the ionization rate {\rc per neutral hydrogen} given by:
\begin{equation}
    \Gamma = \phi\:\!\sigma_{\rm EUV}\exp\left(-\tau\right)
\end{equation}
with $\phi$ the EUV photon flux impinging on the disc. The optical depth ($\tau$) is calculated including both gas and dust to the ionization front:
\begin{equation}
    \tau = \int_\infty^{R_{IF}} n_H\left[\left(1-X_i\right)\sigma_{\rm EUV}+\kappa_{\rm d,EUV}X_dm_h \right]{\rm d}r
\end{equation}
where $R_{IF}$ is the radius of the ionization front. This optical depth is then evaluated by integrating the density profile of the isothermal Parker wind solution out to twenty times the ionization front distance. The velocity structure in this region is given by:
\begin{equation}
u = c_{s,EUV} \sqrt{-W_i\left\{-\left(\frac{R_g}{2r}\right)^4\exp\left[f(r)\right]\right\}}
\end{equation}
where $W_i$ is the $i^{\rm th}$ branch of the Lambert W function (with the 0 branch applicable for the sub-sonic regions and the -1 branch applicable for the supersonic regions \citealt{Cranmer2004}). The function $f(r)$ is given by:
\begin{equation}
    f(r) = 3-\frac{2R_g}{r}
\end{equation}
The density profile can then be obtained through mass conservation:
\begin{equation}
    \rho(r) = \rho_{IF} \frac{u_{IF} R_{IF}^2}{ur^2}
\end{equation}
where $\rho_{IF}$ and $u_{IF}$ are the density and velocity in the EUV heated region at the ionization front, respectively. We define the ionization front to be the radius  at which the ionization fraction drops below $10^{-4}$; however, since the ionization front is very thin, this choice makes no practical difference. For simplicity, we assume a monochromatic spectrum for the EUV field of $\sim$20~eV photons, as such $\sigma_{\rm EUV}=2\times10^{-18}$~cm$^2$. Following \citet{Johnstone1998}, the size of the FUV heated portion is chosen such that its column density provides an optical depth of unity to FUV photons. We then solve this coupled problem such that momentum-flux is conserved through the ionization front. We do this by matching:
\begin{equation}
    \rho_{FUV}\left(R_{IF}\right) c_{s,FUV}^2 = \rho_{IF}\left( c_{s, EUV}^2 + u_{IF}^2\right) \label{eqn:momentum}
\end{equation}
Since the FUV heated region is extremely sub-sonic ($R_d/R_{g, FUV}$ is always $\ll 1$), we have dropped the $\rho u^2$ term in the momentum flux of FUV heated gas into the ionization front. This simplification means we don't need to explicitly solve for mass conservation across the ionization front. Matching the momentum flux across the ionization front is an iterative procedure for which we use the {\sc brentq} solver provided in {\sc scipy} \citep{2020SciPy-NMeth}. The resulting mass-loss rates as a function of disc size are shown in Figure~\ref{fig:photo_rates} for our nominal incident EUV flux of $\phi = 3\times10^{15}$~photons~s$^{-1}$~cm$^{-2}$ and a stellar mass of 1~M$_\odot$.

Unsurprisingly, we find larger mass-loss rates for larger discs and lower dust-to-gas mass ratios. Below a dust-to-gas mass ratio of $X_d\sim 0.03$, the effect of the dust is weak. This weak dependence arises because below $X_d \sim 0.03$ dust has a limited effect on the attenuation of EUV photons; however, it still dominates the attenuation of FUV photons. In EUV photoevaporation, the EUV flux plays the dominant role in setting the mass-loss rates, and the size of the FUV-dominated region plays a minor role. Hence the slow increase in mass-loss rate as the dust-to-mass ratio decreases at low values is due to the slowly increasing size of the FUV-heated region. As the density profile in the FUV heated region is an exponential profile (Equation~\ref{eqn:hydrostatic}), the depth of the region for a given column density (of dust) changes logarithmically. However, above a dust-to-gas ratio of 0.03, the dust absorbs a significant fraction of the EUV flux, resulting in a much lower EUV flux reaching the ionization front. As such, the mass-loss rates fall off rapidly with an increasing dust-to-gas mass ratio. Additionally, we find that mass-loss rates depend on the square root of the ionizing flux (as expected for ionization-recombination equilibrium - \citealt{Johnstone1998}) and exponentially with stellar mass as expected for a sub-sonically launched thermal wind \citep[e.g.][]{Adams2004}. While our mass-loss calculations are 1D and assume spherical divergence, this approach has been shown to give order unity correct mass-loss rates when compared to multi-dimensional simulations \citep{Haworth2019}. Instead of tabulating our mass-loss rates, we choose to fit a physically motivated functional form to the results. This allows us to avoid computationally expensive multi-dimensional interpolation in evolutionary calculations. These functional forms are detailed in the Appendix.

\label{sec:evaporation}

\subsection{Constraints from the observed G-cloud fluxes}\label{sec:brgamma}

One of the clues pointing to the fact the G-clouds are photoionized gas is the detection of the hydrogen and helium recombination lines \citep{Gillessen2012,Burkert2012} as well collisionally excited metal lines in the more recently detected G-clouds \citep{Peissker2020,Ciurlo2020}. Following other authors \citep[e.g.][]{Gillessen2012}, the measured Br-$\gamma$ line luminosities  of order  2--10 $\times 10^{30}$~erg~s$^{-1}$ imply electron densities of order $10^5$--$10^6$ cm$^{-3}$ on scales of $\sim100$~AU. Indeed, as pointed out by \citet{Murray-Clay2012}, if one assumes that these electron densities are outflows at $\sim 10$~km~s$^{-1}$ then one obtains a mass-loss rate of a few $10^{-8}$~M$_\odot$~yr$^{-1}$. This value is roughly the mass-loss rate a photoevaporating disc with a size of a few AU would give you in the Galactic Centre. 

Since our new photoevaporation models explicitly calculate the density and ionization structure of the photoevaporating outflows, we are able to directly compute the expected Br-$\gamma$ luminosity from our models. Assuming case-B recombination, the Br-$\gamma$ luminosity is:
\begin{equation}
    L_{\rm Br\gamma} =  \int  \varepsilon X^2 n_H^2 {\rm d}V \label{eqn:brgamma}
\end{equation}
where $\varepsilon$ is a constant, with a value of $3.44\times10^{-27}$~erg~s$^{-1}$~cm$^{3}$ for $10^4$~K gas \citep[e.g.][]{Ballone2018}. Since our photoevaporating simulations are 1D, we need to turn our density solutions into 3D density structures to compute the line luminosity. Assuming the flow covers the full $4\pi$ solid angle is likely to overestimate luminosity as outflows are launched from the disc's outer edge over a finite solid angle \citep{Richling1998,Richling2000,Adams2004,Haworth2019}. Inspection of the integrand of Equation~\ref{eqn:brgamma} indicates that the contributions from different radii scale as $n^2R^3$. Outside $R_g$, the outflow is likely to expand spherically, inside $R_g$, the outflow is likely to expand super-spherically with an approximately exponential density profile \citep{Owen2021}. This indicates that the Br-$\gamma$ luminosity should be dominated at small radii inside $R_g$. We draw special attention to G2, which, as we discuss in Section~\ref{sec:pressure} is likely to have been pressure confined during its plunge towards pericentre. As such, the argument that the Br-$\gamma$ is dominated at small radii does not hold for G2. For a pressure-confined isothermal structure, the density is approximately constant outside $R_g$ (inspection of Equation~\ref{eqn:hydrostatic} in the limit that $r\gg R_g$, indicates the density will tend to a constant for a hydrostatic profile), and as such the Br-$\gamma$ flux will be dominated at large radii. 

For outflowing density structures, inside $R_g$, the flow is sub-sonic, and the density profile is close to hydrostatic equilibrium. Thus, the density perpendicular to the disc's mid-plane will approximately obey $n_H=n_0\exp(-z^2/2H^2)$ with $H=c_s/\Omega$. This allows us to approximate Equation~\ref{eqn:brgamma} as:
\begin{equation}
    L_{\rm Br\gamma} \approx 2\pi^{3/2} \varepsilon \int X(R)^2 n_0(R)^2 H(R) R {\rm d}R \label{eqn:brgamma_approx}
\end{equation}
where $n_0(R)$ is our density solution to the 1D outflow problem. Finally, we restrict $H(R)$ to be no larger than $2/\sqrt{\pi} R$ which allows an approach to the spherical case in the limit of $H\longrightarrow R$, although this has little impact on the result as $H$ only approaches $R$ outside $R_g$. 
\begin{figure}
    \centering
    \includegraphics[width=\columnwidth]{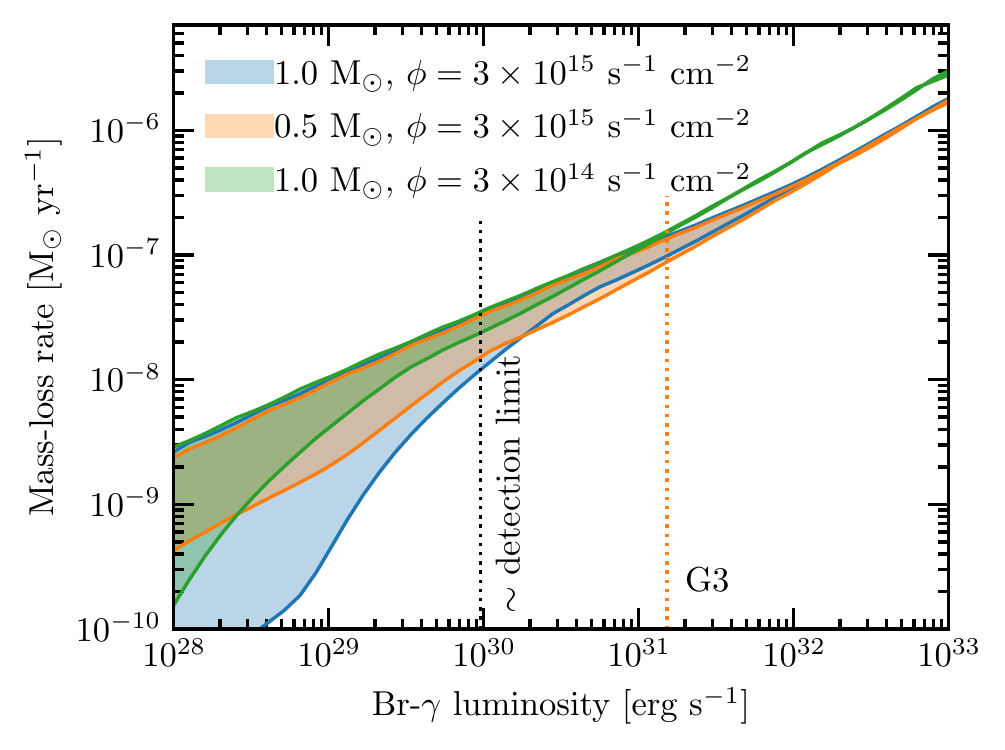}
    \caption{The mass-loss rates from the photoevaporating wind scenario as a function of Br-$\gamma$ line luminosity. The black-dotted line shows the $1\sigma$ error of $2.5\times10^{-4}$~L$_\odot$ from the G2 discovery paper \citep{Gillessen2012}, which is representative of the flux errors reported in other works \citep[e.g.][]{Gillessen2013,Pfuhl2015,Peissker2020}. The orange dotted line shows the most luminous G-cloud, ``G3'' from the discoveries reported by \citet{Ciurlo2020}. Thus, to explain the G-clouds with a photoevaporating disc, one requires mass-loss rates in the approximate range $10^{-8}$--$10^{-7}$~M$_\odot$~yr$^{-1}$.}
    \label{fig:brgamma}
\end{figure}
The results of this calculation are shown in Figure~\ref{fig:brgamma}, where we show the region of parameter space in mass-loss rate and Br-$\gamma$ flux that the disc photoevaporation models can explain.  Comparison between observed Br-$\gamma$ luminosities and the mass-loss rates indicates that if the G-clouds are photoevaporating discs, they are likely losing mass at a rate between $10^{-8}$ and $10^{-7}$~M$_\odot$~yr$^{-1}$. 

Thus, throughout this work, we use the criteria of a photoevaporative mass-loss rate in between $10^{-8}$ and $10^{-7}$~M$_\odot$~yr$^{-1}$ as an indication of matching the observed Br-$\gamma$ line fluxes. As we shall see the evolutionary calculations in our scenarios indicate the discs spend very limited time with a mass-loss rate exceeding $10^{-7}$~M$_\odot$~yr$^{-1}$ and evolve quickly once the mass-loss rate drops below $10^{-8}$~M$_\odot$~yr$^{-1}$, indicating the time our models spend appearing as G-clouds is insensitive to the exact values of these boundaries. 

We also note a difference between our calculations and previous work. In the original consideration of the photoevaporating disc scenario, \citet{Murray-Clay2012} assumed that in the central region, the EUV flux was dominated by the S-stars, and as such in their calculation the EUV flux impinging on the disc rose as G2 approached pericentre, resulting in a Br-$\gamma$ flux that increased by roughly a factor of five. This increase is inconsistent with the observations \citep[e.g.][]{Gillessen2013,Pfuhl2015}. However, as we showed in Section~\ref{sec:radiation} (Figure~\ref{fig:EUV_flux} we find the EUV-flux is dominated by the O/WR disc and it remains approximately constant as one approaches Sgr A*. This means the predicted Br-$\gamma$ fluxes from the photoevaporating disc scenario would not actually strongly rise as inferred by \citet{Murray-Clay2012}, but would rather remain roughly constant or slightly increase. Although, we note as mentioned above (and discussed in \ref{sec:pressure}), we suspect G2 is actually a pressure-confined photoevaporating flow throughout its observational history. This fact means it's not trivial to map our photoevaporating wind Br-$\gamma$ fluxes directly onto G2's. 

\subsubsection{NIR luminosities}\label{sec:NIR}

\begin{figure}
    \centering
    \includegraphics[width=\columnwidth]{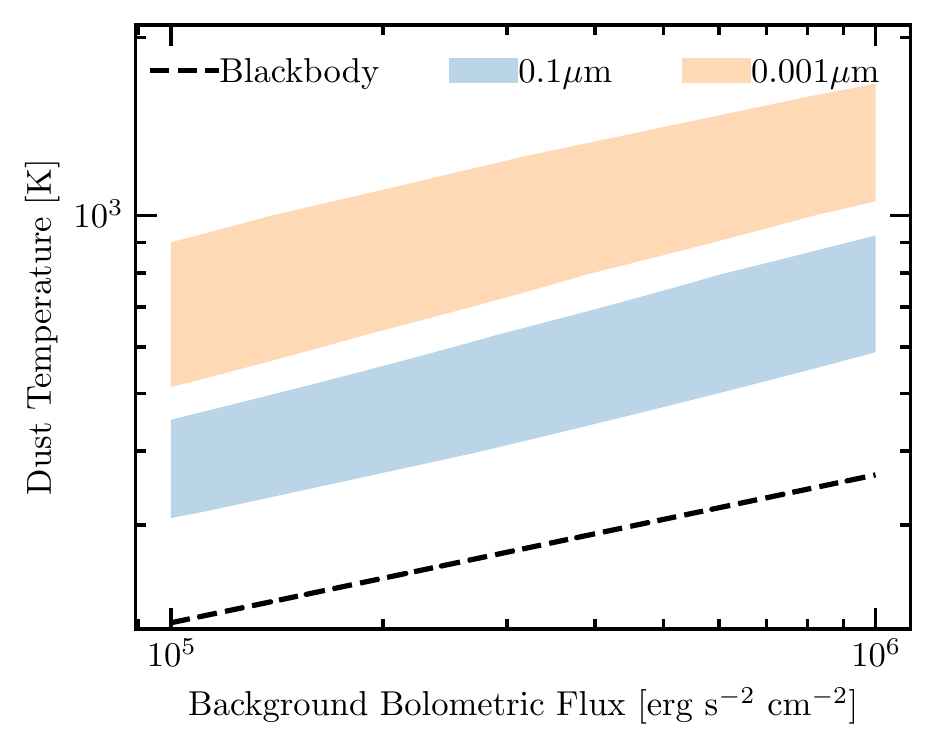}
    \caption{The temperature of 0.001 and 0.1 $\mu$m size silicate grains in the Galactic Centre as a function of bolometric flux, in addition to the blackbody temperature. The bands represent ranges arising from the plausible range in the temperature of the radiation field, as well as the range of solid angles over which absorption could take place (between $4 \pi$ for an isotropic radiation field and $\pi$ of a plane-parallel radiation field). This demonstrates that it is possible that the dust in the G-clouds is heated by the external radiation field to temperatures $\sim$600~K. }
    \label{fig:dust_temp}
\end{figure}

The observed NIR luminosities of the G-clouds are less constraining. \citet{Zajacek2017} used radiative transfer modelling to show that the NIR properties of the G-clouds were consistent with those of a young T Tauri star surrounded by a disc. The emission demonstrates the presence of dust, but it's still currently unclear what source(s) are heating the dust.  In the T Tauri star scenario, it could either be the young star itself, the combined background emission from all the nearby stars, or both.  {\rc We can investigate this more explicitly using our analysis from Section~\ref{sec:radiation}. During our computation of the typical EUV fluxes, we also compute the properties of the bolometric radiation field, finding fluxes in the range $10^5-10^6$~erg~s$^{-1}$~cm$^{-2}$, with mean temperatures $\sim 2-4\times10^4$~K. Using these values we can compute the temperatures of optically thin dust particles, using the optical constants for astronomical silicate grains \citep{Weingartner2001}. The results are shown in Figure~\ref{fig:dust_temp}, where small sub-micron particle sizes can be heated well above the black-body temperature to temperatures $\sim$600~K due to their efficient absorption in the UV and inefficient emission in the IR. However, without knowledge of the dust size distribution in the G-clouds, it's not possible to conclude whether the dust is externally or internally heated.}
 However, we can place a few constraints on the requirement that there is sufficient dust to match the required luminosities. The most constrainable scenario we present in this work is where we explain the G-clouds as a dust-depleted photoevaporating disc, and therefore we should check we have sufficient dust mass to explain the observed NIR luminosities. The K band luminosity for G2 is debated. However, there is little debate at longer wavelengths with G-clouds typically presenting a $\sim 600~$K black-body-like spectrum with a luminosity of a few Solar luminosities, corresponding to a black body emitting radius of $\sim 1$~AU \citep[e.g.][]{Gillessen2012,Peissker2020}.  Such a temperature and size rule out black-body emission from an optically thick disc, internally heated by a central star, as it is not possible to reach such high temperatures at such large distances without violating the total luminosity constraints. Optically thin, small dust particles that are heated above their blackbody temperature due to their inefficient emission of IR radiation are more likely (Figure~\ref{fig:dust_temp}). Taking the power emitted by a single dust grain to be $4\pi a^2 Q \sigma T^4$, where $Q$ is the radiative efficiency, which we approximate to be $Q\sim 2\pi a T / b_w$ (with $b_w$ Wein's displacement constant),  we find the total mass in dust must be:
\begin{equation}
    M_{\rm dust} \sim 1\times10^{-3}~{\rm M}_\oplus~\left(\frac{L_G}{1~{\rm L}_\odot}\right)\left(\frac{T}{600~{\rm K}}\right)^{-5}\left(\frac{\rho_{\rm in}}{1~{\rm g~cm}^{-3}}\right) 
\end{equation}
Where we have inserted representative values for the G-clouds NIR luminosities ($L_G$), temperatures, and the internal density of the dust particles ($\rho_{\rm in}$). This mass of dust would become optically thick to external irradiation if it had a size smaller than:
\begin{equation}
    R_{\tau=1}\approx 8~AU \left(\frac{M_{\rm dust}}{10^{-3}~{\rm M}_\oplus}\right)^{1/2}\left(\frac{\kappa_d}{10^4~{\rm cm^2~g^{-1}}}\right)^{1/2}
\end{equation}
where we have assumed that the heating source is UV photons from the nearby massive stars, adopting an opacity of $10^4$~cm$^2$~g$^{-1}$ \citep[e.g.][]{Ali2021}; for cooler irradiating sources (e.g. an internal low-mass star) the opacity will be lower and size-scale limit correspondingly smaller.  Comparisons to our outflow solutions indicate that they contain sufficient mass, and have large enough length scales that they can satisfy the above mass and optical depth requirements unless the disc is small ($\lesssim 3$~AU) and are significantly depleted in dust $X_d \lesssim 10^{-4}$, a region rarely traversed by our evolutionary models. 

\section{Standard Evolution Scenarios}\label{sec:standard}
Our mass-loss rates, calculated in Section~\ref{sec:evaporation} provide a fundamental problem to the protoplanetary disc origin of the G-clouds. Specifically, it is difficult to make a protoplanetary disc, born at a similar time to the young cluster, live sufficiently long while providing a large enough mass-loss rate at late times to be observable as a G-cloud.  

Figure~\ref{fig:photo_rates} shows the rough range of photoevaporation rates consistent with the observed Br$\gamma$ fluxes for the G-clouds. The solid lines show contours of a constant mass-loss timescale of 3~Myr. As discussed in \citet{Murray-Clay2012} a massive disc ($\sim 0.1$~M$_\odot$) would have a mass-loss timescale of a few Myr for a disc with the standard dust-to-gas ratio of 0.01 and sizes of $\sim 10$~AU, while being compatible with the observed Br$\gamma$ flux of G2. While the above argument is indicative that one might be able to explain the G-clouds properties, it neglects that photoevaporation causes the disc to evolve and accrete rapidly \citep[e.g.][]{Clarke2007,Rosotti2017, Winter2020}. As photoevaporation removes material from the outer disc, reducing its radius, the viscous transport time of the disc decreases ($t_\nu$ approximately scales linearly with radius for a constant $\alpha$, passively heated disc -- \citealt{Hartmann1998}). Shorter viscous transport timescales allow the disc to lose angular momentum (by providing it to the photoevaporating wind) more efficiently, allowing the disc to accrete more efficiently. In Figure~\ref{fig:photo_rates}, we show the disc radius (line labelled $R_{\rm disc}$) photoevaporation would reach by removing material from a 0.1~M$_\odot$ disc with a $\Sigma\propto R^{-1}$ profile and initial radius of 20~AU in 3~Myr as the dot-dashed line (significantly bigger discs would be tidally stripped by the black hole's tidal field). In addition, we show the radius where the viscous transport time of the disc is equal to 1~Myr (for $\alpha = 10^{-3}$). These two lines allow us to sketch an evolutionary pathway: for an initially large disc, photoevaporation will strip the disc, reducing its radius. Once the disc is photoevaporated to the point where the viscous transport time is of order its age, accretion takes over, and the disc rapidly accretes onto the star. This pathway is confirmed in our numerical calculations shown later. The final accretion stage takes place at approximately fixed radii, where the outward transport of angular momentum in the disc is balanced by the removal of angular momentum in the wind \citep[e.g.][]{Winter2020, Owen2021}.  Thus, accretion acts to shorten the disc's lifetime from what would be obtained by estimating instantaneous mass-loss timescales (the disc mass depletes as $\propto \exp(-t/t_{\nu})$, rather than a power-law fashion for discs with a fixed outer radius \citealt{Rosotti2018}). Figure~\ref{fig:photo_rates} indicates that it is difficult to simultaneously have a protoplanetary disc survive for at least a few Myr while providing sufficient mass-loss rates at late times to explain the Br$\gamma$ fluxes observed for the G-clouds. Indeed, these issues are similar to the ``propyld lifetime'' problem identified in more benign star-forming regions \citep[e.g.][]{Henney1999,Haworth2022}. However, the proposed solution for the Orion Nebula Cluster of dynamical evolution, coupled with an extended period of star-formation \citep[e.g.][]{Winter2019}, is unlikely to hold in the Galactic Centre. This is because the EUV fluxes are still large at distances of $\gtrsim1$~pc, so it is difficult to argue that recently formed stars have just been scattered in.  

In the following sub-sections, we demonstrate these insights explicitly through direct numerical calculation of the disc evolution. In the following sections, we modify the standard scenario to arrive at an evolutionary picture that allows a protoplanetary disc to survive for a few Myr in the Galactic Centre and provides sufficient mass-loss rates at late times to explain the observed G-clouds. We also comment on the evolution of any circumstellar disc formed recently, perhaps through a stellar merger. 

\subsection{Numerical Models}\label{sec:numerical}

To model the evolution of a photoevaporating circumstellar disc in the Galactic Centre, we numerically solve the viscous thin disc model for both the gas and dust components. In this formalism, the surface density of the gas, $\Sigma_g$, and dust $\Sigma_d$, evolve according to:
\begin{equation}
\pderiv{\Sigma_g}{t} = \frac{1}{R} \pderiv{}{R}\left[3 R^{1/2}\pderiv{}{R}\left(\nu \Sigma_g R^{1/2}\right) - \frac{2 \Lambda \Sigma_g}{\Omega} \right] - \dot{\Sigma}_{w,g},
\end{equation}
and 
\begin{equation}
\pderiv{\Sigma_d}{t} = - \frac{1}{R} \pderiv{}{R}\left[R v_d \Sigma_d - \frac{\nu}{Sc} R \Sigma_g \pderiv{}{R}\left(\frac{\Sigma_d}{\Sigma_g}\right) \right] - \dot{\Sigma}_{w,d},
\end{equation}
where $R$ is the cylindrical radius, $t$ is the time, $\nu  = \alpha c_s H$ is the kinematic viscosity, $\Omega$ is the Keplerian angular frequency, $v_d$ is the velocity of the dust, $Sc$ is the dust's Schmidt number which we set to unity, $\Lambda$ is any external torque on the disc (e.g. from a planet) and $\dot{\Sigma}_w$ is the photoevaporation rate of gas (sub-script $g$) and dust (sub-script $d$). This model is numerically integrated using the method outlined in \citep{Owen2014,Booth2020}, which is first-order in time and second-order in space and uses a van-Leer limiter for advection terms. We use a grid extending from 0.025~AU to 40~AU with 500 non-uniformly spaced cells (the spacing follows $R^{1/3}$, yielding higher resolution at smaller radii). At the outer boundary, we assume that the tidal field removes material instantly from the disc and use outflow boundary conditions (in all cases, this doesn't matter as photoevaporation truncates the disc inside the inner boundary). The dust-size distribution is evolved using the formalism of \citet{Birnstiel2012}, using a fragmentation velocity of 10~m~s$^{-1}$. This dust-size evolution accounts for growth due to sticking and fragmentation due to destructive collisions, we assume that 10\% of collisions below the fragmentation velocity result in growth \citep{Booth2020}. We note that these choices for the fragmentation velocity and fraction of collisions that result in growth do not change the general picture of the evolutionary results shown below. The external photoevaporation rates from our calculation described in section \ref{sec:evaporation} are implemented using a formalism similar to \citet{Clarke2007}. Our procedure is as follows: we define the disc radius to be the point in the disc where the gas surface density drops below $10^{-4}$~g~cm$^{-2}$. We then measure the dust-to-gas ratio by considering the amount of dust and gas that resides within 10 cells of the disc radius to estimate the dust-to-gas mass ratio in the photoevaporative outflow. We then remove material from cells, moving inwards until we reach the required total mass-loss. Cells that have material completely removed are assigned a surface density of the floor value of 10$^{-20}$~g~cm$^{-2}$. The calculation of the dust-to-gas ratio over a radial range of 10 cells at the outer boundary is necessary as the finite nature of the grid can change the exact value of the pressure gradient (and hence the dust-to-gas ratio) in the last cell to change by large amounts from time-step to time-step. Using just the last cell to measure the dust-to-gas mass ratio causes it to fluctuate, and hence the mass-loss rate to fluctuate strongly, causing the disc to occasionally nonphysically expand and then contract. By experimentation, we find that 10 cells provide a sensible balance between smoothing the calculated dust-to-gas ratio and using a small radial range near the outer edge. We note that in these 10 cells, the dust advection is directed outwards. We find the exact choice of the number of cells to measure the dust-to-gas mass ratio over or the threshold surface density to define the disc's outer edge does not change the general results of our calculations. 

\subsection{Results: protoplanetary disc lifetimes in the Galactic Centre}

In order to directly verify our insights that a protoplanetary disc is unlikely to survive the $\gtrsim 3$~Myr required to appear as G-cloud today, we simulate the evolution of a disc receiving a  photon flux of $3\times10^{15}$~photons~s$^{-1}$~cm$^{-2}$. We make a number of choices to maximise the disc lifetimes. The luminosity constraints on G2 of $\lesssim 5$L$_\odot$ \citep{Gillessen2012} implies that if the central source were a young star, it would have to be less massive than $\sim 1.4$~M$_\odot$. As more massive stars have deeper gravitational potential wells, making them harder to photoevaporate, we chose our central star to have a mass of 1.4~M$_\odot$. We assume an initially compact disc (such that initial tidal removal of material is minimal, maximising its mass), where the initial profiles follow a Lynden-Bell \& Pringle zero-time similarity solution \citep{LyndenBell1974}:
\begin{equation}
    \Sigma_g = \Sigma_{0} \left(\frac{R}{R_0}\right)^{-1}\exp\left(-\frac{R}{R_0}\right)
\end{equation}
The initial disc mass is set to be 0.15 of the stellar mass, $R_0$ is set to 5~AU, the dust-to-gas mass ratio is assumed to be uniform everywhere with a value of 0.01, and the dust grains are taken to initially be 0.1$\mu$m in size. The disc's mid-plane temperature structure follows a $T\propto R^{-1/2}$ profile, appropriate for a disc passively heated by the central star \citep[e.g.][]{Kenyon1987}. The disc temperature is set to 250~K at 1 AU, appropriate for a flaring, passively heated disc around a 1.4~M$_\odot$ star. While there is debate as to whether the external radiation field contributes to heating the dust in G-clouds \citep[Section~\ref{sec:NIR}, e.g.][]{Gillessen2012,Peissker2020}, a lower disc temperature causes the disc to evolve slower, extending its lifetime.  Since we are applying no external processes to the disc in this first set of models, there is no external torque ($\Lambda=0$). In these calculations, the standard zero-torque boundary condition is applied at the inner boundary \citep[e.g.][]{Pringle1981,Pringle1986}.

\begin{figure}
    \centering
    \includegraphics[width=\columnwidth]{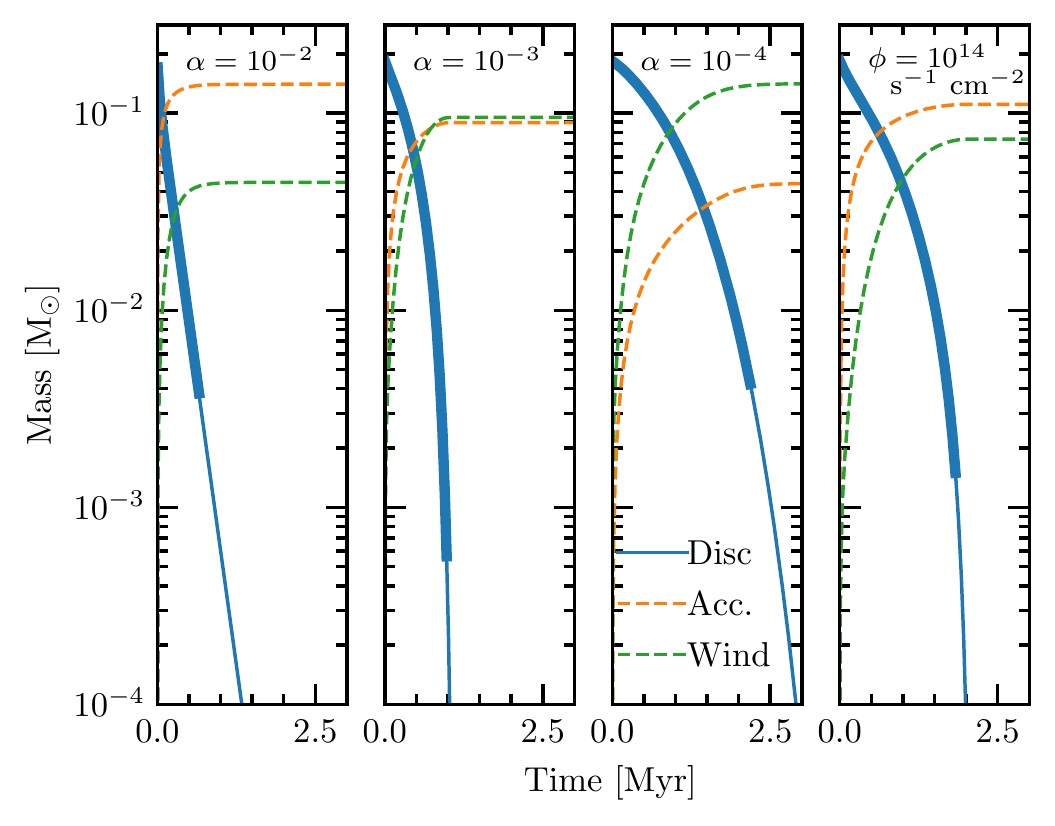}
    \caption{The evolution of the disc mass and the total masses lost through accretion and in the photoevaporative wind. The thick portion of the disc mass lines shows when the photoevaporative rate exceeds $10^{-8}$~M$_\odot$~yr$^{-1}$ (and is visible as a G-cloud in Br-$\gamma$). The first three panels show different values of the turbulent viscosity for a standard EUV photon flux of $3\times 10^{15}$~photons~s$^{-1}$~cm$^{-2}$, the final panel shows a low EUV scenario (i.e residing in the outer regions of the Young disc) with an $\alpha=10^{-3}$. It is extremely difficult to have a disc survive for $\gtrsim 2$~Myr with photoevaporation rates sufficient to explain the G-clouds.  }
    \label{fig:standard_lifetime}
\end{figure}

The resulting evolutionary tracks for a range of plausible parameters are shown in Figure~\ref{fig:standard_lifetime}. All these show disc lifetimes considerably shorter than the required value of $\gtrsim 3$~Myr, and in the more nominal cases have sufficiently small photoevaporation rates that they would not be observed as G-clouds beyond 1~Myr. 

These results, appear at first glance to not match the calculations that appear to indicate photoevaporation will only deplete discs of 5-10~AU in scale on timescales of several Myr \citep{Murray-Clay2012}. However, as discussed above, these calculations only focused on the removal of disc material by photoevaporation. Whereas, the numerical calculations show that disc accretion onto the central star is an equally, if not more powerful disc depletion mechanism.  Thus, through accretion, photoevaporation brings about a rapid end to the disc's life. Even in the cases of extremely low viscosities ($\alpha=10^{-4}$) or low photon-fluxes ($10^{14}$~photons~s$^{-1}$~cm$^{-2}$) with favourable choices in the initial conditions we cannot make discs survive around young stars sufficiently long to provide the source population of G-clouds unless we somehow modify the disc's evolutionary pathway. 

\section{Instantaneous formation of a disc}

One trivial solution to the disc lifetime problem would be to suppose that the disc does not surround a young star born in the star formation episode $\sim 3$~Myr ago, but rather the disc has recently formed. This disc could have formed either as the by-product of a stellar merger \citep[e.g.][]{Prodan2015,Naoz2016}, or through interactions with compact objects (such as black holes), as suggested by \citet{MiraldaEscude2012}. In the following discussions, we {\it assume} that it is the disc, that creates the G-cloud signature\footnote{It could be that this collisional byproduct itself is responsible for the G-cloud signature; however, since the properties of these byproducts are uncertain, we focus our discussion on any disc created.}. In this case, if the formed disc had initial conditions similar to those chosen in the previous section, i.e. similar to discs formed through the star formation processes then there is no lifetime issue. However, the discs used in the previous calculations are too extended and massive to have formed recently through such processes. Thus, here we consider more compact, lower-mass discs. 

If the G-clouds were to have formed recently, they would be in similar orbits to where we see them today, with most spending some time in their orbit inside 0.01~pc. The tidal radius at separations $\lesssim 0.01$~pc is now only a few AU. Thus, we choose to set our initial disc scale radius for these calculations to be $R_0=1$~AU. The initial disc masses are more uncertain, but \citet{MiraldaEscude2012} appeal to planetary collision simulations \citep[e.g.][]{Canup2004} which suggest disc masses of $M_d/M_* \lesssim 10^{-2}$. 

\begin{figure}
    \centering
    \includegraphics[width=\columnwidth]{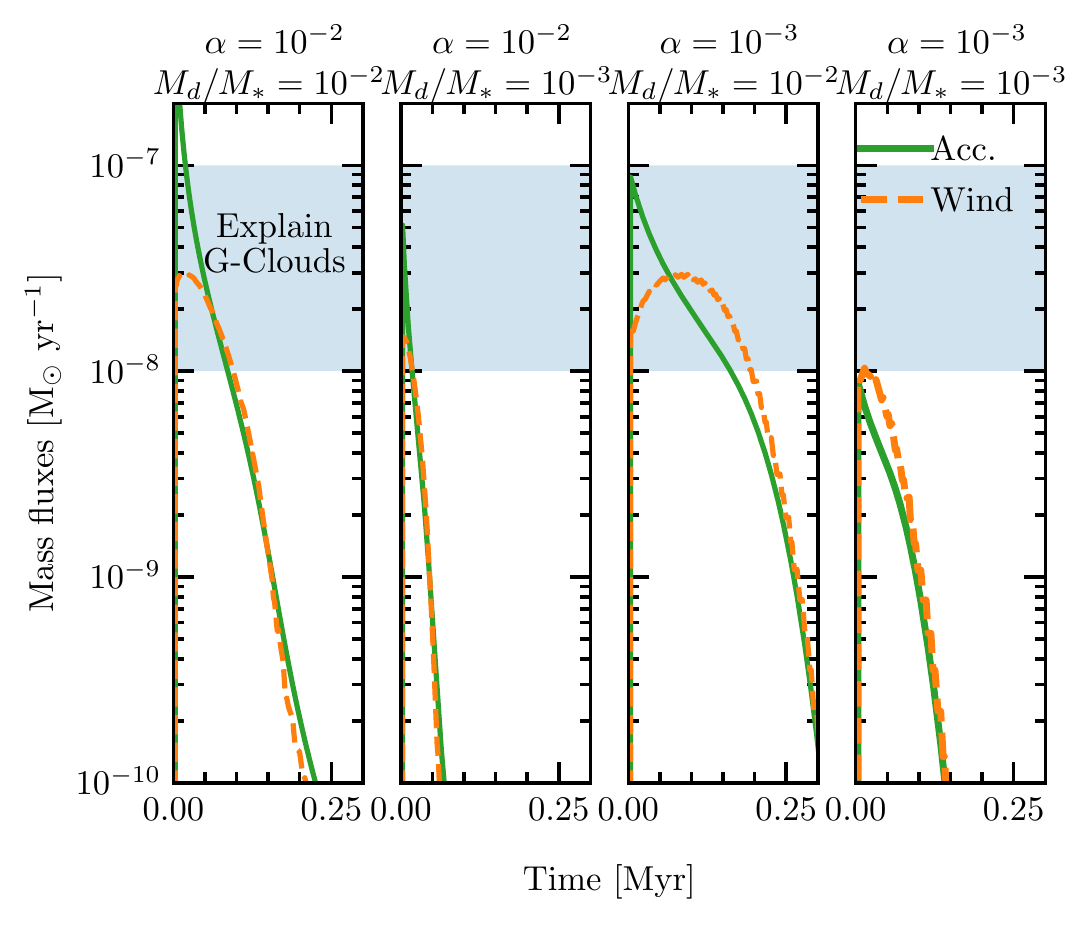}
    \caption{The mass-accretion (solid) and photoevaporation rates (dashed) for a low-mass, compact disc that has been recently created.  When the photoevaporation rate sits in the shaded region the photoevaporative flow is sufficiently dense to explain the observed Br-$\gamma$ luminosities of the G-clouds. Each panel shows a different evolutionary calculation where the initial disc mass and viscous $\alpha$ parameter has been varied. }
    \label{fig:instant_disc}
\end{figure}

The resultant evolutionary calculations, shown in Figure~\ref{fig:instant_disc} show that these models can explain the lower range of mass-loss rates $\lesssim 3\times10^{-8}$~M$_\oplus$~yr$^{-1}$ for short periods of time $\lesssim 0.1$~Myr. However, reproducing such cases requires massive initial discs and some amount of fine-tuning for the value of the viscosity. Even so, we do struggle to reproduce any model with sufficiently high photoevaporation rates to explain the Br-$\gamma$ fluxes of the more luminous G-clouds (e.g. G3 and G4 \citealt{Ciurlo2020}). Thus, while this scenario could work to explain the G-clouds, the short disc-lifetime of $\lesssim 10^5$~years, would require a production rate of $\gtrsim 1/10^{4}$~yr$^{-1}$ to explain the fact we current observe $\sim 5-10$ objects. This exceeds the rate estimated by \citet{MiraldaEscude2012} for interactions between stars and compact objects. Alternatively, in a merger scenario, the rate would imply that $\sim 300$ mergers need to have occurred since the last star-formation episode. Since these young stars which formed in binaries are the most likely progenitors for such mergers \citep{Ciurlo2020}, as binaries do not survive long in the dense environment of the Galactic Center \citep[e.g.][]{Stephan2019}. \citet{Stephan2016} estimated that $\sim 10$\% of binaries would merge in the Galactic Center within a few Myr. Thus, one would require a population of $\gtrsim 3000$ young, low-mass binaries to have formed in the last star formation episode. Additionally, \citet{Stephan2016} estimate that only 10\% of Solar-like stars can exist in binaries in the Galatic Center (as opposed to the 50\% in the field). Thus, one would require a population of $\gtrsim 30,000$ young T-Tauri stars to have formed in the star-formation episode. This estimate exceeds the $\sim$2000 T-Tauri stars estimated to have formed based on extrapolations of the observed IMFs, and X-ray surface brightness constraints  (e.g. \citealt{Lu2013}, see Section~\ref{sec:rates}). Even if all the young binaries could be made to merge within a few Myr, one could still not explain the number of observed G-clouds through the instantaneous formation of an accretion disc.     

\label{sec:instant}

\section{Non accreting Disc}\label{sec:non_acc}

The problem identified in the previous Sections is photoevaporation enhances accretion onto the star. 
Therefore, if accretion could be prevented, this would provide a promising avenue for extending a disc's lifetime. 

Young, low-mass stars ($M_* \lesssim 1.5$~M$_\odot$) are known to accrete through magnetospheric accretion \citep{Bouvier2007,Hartmann}, where the strong magnetic field of the young star ($B_*\sim 1$kG) truncates the disc, funnelling material via accretion streams onto the stellar poles. This truncation radii occurs when the magnetic torque from the star's magnetic field is sufficiently strong to overcome pressure (thermal and ram) of material in the disc, disrupting it \citep[e.g.][]{GoshLamb1,Koenigl1991}. In general, this occurs at a radius where the orbital frequency of the disc is faster than the star's rotation period. However, if the star is spinning sufficiently rapidly, the system can enter a ``propeller'' regime \citep[e.g.][]{Lovelace1999,Romanova2003}. In this case, the truncation occurs outside the co-rotation point, now, material at the inner edge of the disc has a rotation velocity slower than the stellar magnetosphere. In this case, the disc material is prevented from accreting, and the star provides a torque to the disc causing it to expand, resulting in a  ``decretion'' disc \citep{Pringle1991}, and in some cases, a powerful wind from the inner regions can be launched \citep[e.g.][]{Romanova2004}. 

In regular T Tauri stars, this process happens toward the end of the disc's lifetime when the accretion rate and disc mass are low \citep[e.g.][]{Armitage1996}. However, there is ample reason to think it might happen earlier in the Galactic Centre at higher disc masses. In the standard picture, it is the accretion of disc material that modulates the stellar spin, maintaining the disc inner edge close to co-rotation, allowing accretion, a process known as disc locking \citep[e.g.][]{Shu1994,Hartmann2002}. However, it has been recently hypothesised that the removal of disc material, through photoevaporation, stops the star from accreting as much material as normal, preventing the slowing of its rotation. In addition, in the Galactic Centre, the disc's mass is lower at an earlier age, so when the young T Tauri star begins to contract and spin up, there is less disc material to break the star's rotation. Thus, strong external photoevaporation produces a population of T Tauri stars that are rapidly spinning (periods $\sim$ 1~day, rather than the more typical periods of $\sim$ 10~days, \citealt{Lee2017}) by the age of $\sim 1$~Myr \citep{Roquette2021}. Indeed, there is observational support for this concept, where T Tauri stars in more massive clusters (which produce higher UV fields, and hence stronger external photoevaporation) are more rapidly rotating \citep{Roquette2021}. 

With this in mind, we can estimate the accretion rate (and hence approximate disc mass) at which a rapidly spinning T Tauri star would prevent accretion and transition from an accretion disc to an outflow-dominated decretion disc. The stellar magnetic field truncates a disc at a radius of approximately \citep[e.g.][]{GoshLamb2}:
\begin{equation}
    R_T\approx R_* \left(\frac{B_*^4R_*^5}{GM_*\dot{M}_*^2}\right)^{1/7}
\end{equation}
requiring that the truncation radii is at a larger radius than the radius of co-rotation, yields:
\begin{equation}
    \dot{M}_* < \frac{B_*^2R_*^6\Omega_*^{7/3}}{\left(GM_*\right)^{5/3}}
\end{equation}
or, putting in typical values:
\begin{eqnarray}
   \dot{M}_* &\lesssim& 5.9\times10^{-9}~{\rm M}_\odot~{\rm yr}^{-1} \left(\frac{B_*}{1~{\rm kG}}\right)^2\left(\frac{R_*}{1.5~{\rm R}_\odot}\right)^6\nonumber \\&\times&\left(\frac{M_*}{1.4~{\rm M}_\odot}\right)^{-5/3} \left(\frac{P}{3~{\rm days}}\right)^{-7/3}  \label{eqn:mdot_crit}
\end{eqnarray}
This mass accretion rate yields a disc mass of order $10^{-3}$~M$_\odot$ inside 10~AU for a viscous alpha value of $10^{-3}$, consistent with our requirements from Figure~\ref{fig:photo_rates}. At this mass-accretion rate, the angular momentum transfer rate between the star and disc $\dot{L}\sim \dot{M}_*\sqrt{GM_*R_T}$ can be compared to the angular momentum in the star ($L=I_*\Omega_*$) to determine the spin-down timescale of the star. Evaluating for the same representative parameters above yields a spin-down timescale of $\sim 5$~Myr. This spin-down timescale is longer than, or comparable to the spin-up timescale due to the pre-main-sequence contraction of the star (a few Myr). This means that even though the star is transferring angular momentum to the disc, the star's contraction still means it will spin up, allowing accretion to be halted indefinitely. However, if the star were to have a slower spin rate, the reverse would be true and stellar contraction would be unable to overcome the spin-down torque on the disc, the star's spin would slow, and accretion would proceed. Indeed, in Section~\ref{sec:param}, we find a critical value of the star's initial spin to be around 4-5 days, with spin periods less than this value suppressing accretion at disc masses of order $10^{-2}-10^{-3}$~M$_\odot$ and slower spins allowing accretion to continue to very low disc masses.

\begin{figure}
    \centering
    \includegraphics[width=\columnwidth]{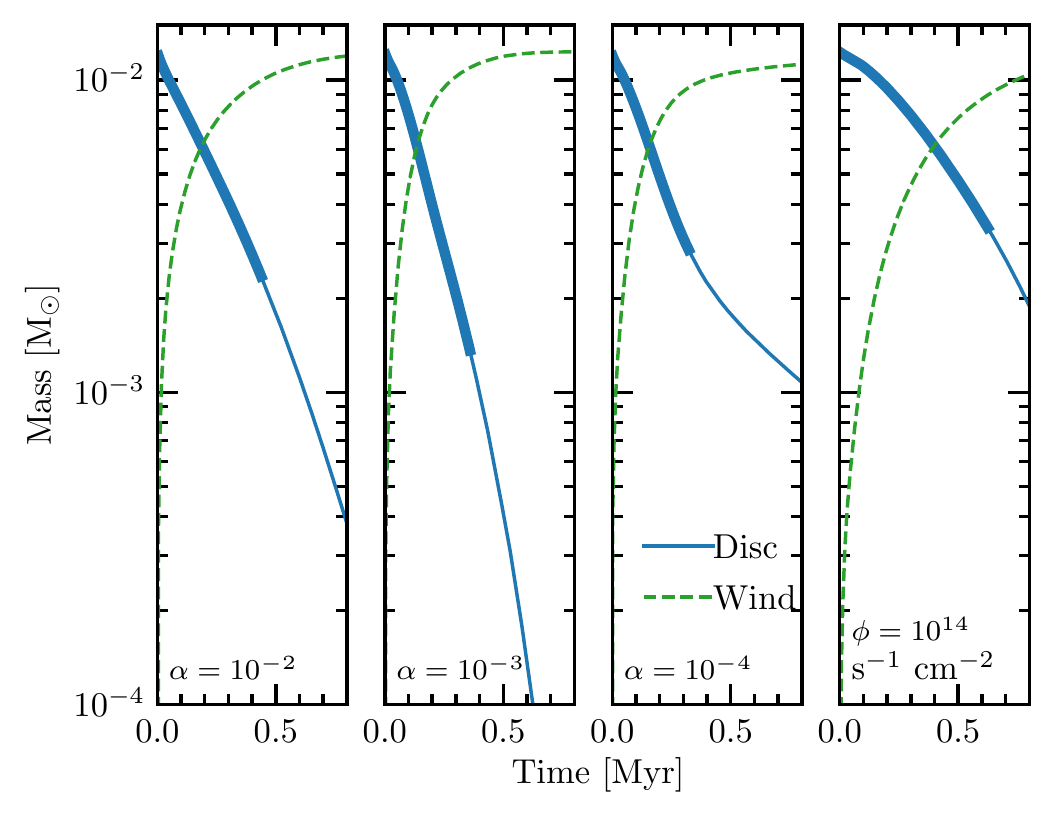}
    \caption{The evolution of the disc mass and the total masses lost in the photoevaporative wind, in the case that the disc is prevented from accreting. The thick portion of the disc masses shows when the photoevaporative rate exceeds $10^{-8}$~M$_\odot$~yr$^{-1}$. The first three panels show different values of the turbulent viscosity for a standard EUV photon flux of $3\times 10^{15}$~s$^{-1}$, and the final panel shows a low EUV scenario (i.e residing in the outer regions of the Young disc) with an $\alpha=10^{-3}$. It is extremely difficult to have a disc survive for $\gtrsim 2$~Myr with photoevaporation rates sufficient to explain the G-clouds.  }
    \label{fig:no_acc}
\end{figure}

In order to assess whether this effect can shut off accretion and provide sufficiently long disc lifetimes, we model the evolution of our discs as in Section~\ref{sec:numerical}, but this time we set the initial disc mass to $0.01$~M$_*$. This choice of initial condition represents an optimistic upper limit for the remaining disc material at the point that accretion shuts down. In order to prevent accretion in our numerical scheme, the inner boundary is given a zero-flux boundary condition, causing the disc to behave like a decretion disc. In this case, the torque from the star causes the disc to expand more rapidly, enhancing the photoevaporation rate as the material is driven outwards by the disc, where it is easier to escape. The resulting disc lifetimes (after accretion ceases) are shown in Figure~\ref{fig:no_acc}. These show that preventing accretion alone cannot be the answer, as the enhanced photoevaporation now destroys the discs fairly rapidly $\lesssim 0.5$~Myr. Thus, it would require considerable amounts of fine-tuning for the observed G-clouds to be in the phase where accretion has just ceased, and we're witnessing their final destruction. 

However, should the disc have just formed via a merger \citep[e.g.][]{Prodan2015} (see discussion in Section~\ref{sec:instant}), then the prevention of accretion by a strong magnetic field could allow the disc to survive sufficiently long to explain the G-cloud population. We speculate here that such a merger would have to create a rapidly spinning, highly magnetised central object. The prevention of accretion in the merger scenario solves several of the issues posited in the previous section. Namely, we can now reach the higher mass-loss rates $\sim 10^{-7}~$M$_\odot$~yr$^{-1}$ required to explain the more luminous G-clouds. In addition, the longer disc lifetimes allow the discs to be observed as G-clouds for $\sim 4\times10^5$~yrs. In this case, we would require roughly 75 mergers to have taken place in the last 3~Myr, at a formation rate of $\sim 2.5\times10^{-5}$~yr$^{-1}$, which is still higher than the rate from \citet{MiraldaEscude2012}. Again following \citet{Stephan2016}, who demonstrated that $10\%$ of stars could be in binaries in the Galactic Centre, then our 2000 Tauri stars could yield $\sim 200$ low-mass binaries. A 10\% merger fraction within a few Myr would again fail to explain the number of observed G-clouds. However, unlike the case discussed in Section~\ref{sec:instant}, there is at least a sufficient number of progenitor binaries in this scenario. Thus, we speculate it may be possible to increase the merger rate. For example, through the use of evection resonances, where even co-planar binaries can be excited to high eccentricity, this mechanism has been suggested as a process to enhance the rate of binary black-hole mergers \citep{Bhaskar2022}.

This merger scenario should be considered in the future to determine whether the binary origin for G-clouds can match the requirements in disc mass and create a rapidly spinning magnetised central object with sufficient frequency. We shall not do such calculations here and instead suggest an alternative scenario that naturally arises out of the formation and evolution of young T-Tauri stars in the following section.

\section{Shutdown of accretion and photoevaporation}

To make a circumstellar disc around a T Tauri star survive for several Myr, we both need to shut down accretion and photoevaporative mass-loss. While this might seem like a difficult proposition, there is a natural solution. As already discussed in Section~\ref{sec:evaporation} the photoevaporation rate can be significantly reduced if the gas becomes enhanced in dust. Furthermore, as discussed in the previous section, T Tauri stars in the Galactic Centre are likely to be rapid rotators, preventing accretion at disc masses of order $\sim 10^{-3}$~M$_*$, forcing the disc into the propeller regime at higher disc masses that standard T Tauri stars. 

There is a well-known sub-population of protoplanetary discs that appear to have high dust-to-gas mass ratios in their outer regions: ``transition discs'' \citep{Andrews2011,vanderMarel2015,vanderMarel2016}. In these discs, the gas contains a pressure maximum that traps dust preventing it from drifting towards the star, resulting in high, localised dust-to-gas ratios that can approach unity \citep{Bruderer2014,Ubeira2019}. This trapping of dust has been confirmed observationally, where the material actively accreting onto the star is observed to be depleted in refractory elements \citep{Kama2015}. While it is still not entirely clear what causes the pressure maximum \citep{Owen2016}; a favoured explanation is a planet which is sufficiently massive to open a gap in the gas disc, resulting in a pressure maximum exterior to its orbit \citep{Rice2006,Zhu2012,Pinilla2012}. This picture matches the disc morphology and architecture in PDS 70 system \citep{Keppler2019}. 

Therefore, we consider a final scenario where the disc forms a giant planet at a few AU and follow the combined evolution of the star, disc, and planet. As indicated above, the disc switches from an accretion disc to a decretion disc once the disc becomes sufficiently depleted, and the star is spinning rapidly. In order to capture this process, we model the combined interaction between the magnetic star and the disc. To do this, we follow the formalism laid out in \citet{Pringle1992} and \citet{Armitage1996} for the interaction between the stellar magnetic field and disc. This allows us to insert a non-zero torque $\Lambda$ in the inner disc, allowing the disc to exist in either the accretion or propeller regime and switch between the two based on the disc's properties. This star-disc coupling also spins up or spins down the star, and we calculate the evolution of the star's spin in an identical way to \citet{Armitage1996}. We take the star's radius to evolve in an identical manner to \citet{Armitage1996}, which itself is a fit to stellar evolution models from \citet{Pols1995}. Since we are focused on the early evolution of the star on timescales less than ($\lesssim 10$~Myr), shorter than the timescales for typical Sun-like stars to form radiative cores, we fix the star's moment of inertia to $0.2 M_*R_*^2$. The planet's effect on the disc is included as another torque, arising from the impulse approximation \citep{Lin1986} and is implemented, including migration, using the approach of \citep{Alexander2012}, as done by \citet{Booth2020}. One critical issue in 1D simulations of planet-disc interactions is how much material flows across the planetary gap. This flow must be parameterised and is uncertain even in the case of regular disc evolution. We use the approach from \citet{Alexander2009} where the steady-state viscous mass-flux (i.e. $3\pi\nu\Sigma$) is measured outside the planet's orbit. A fraction of this mass-flux $f_{\dot{M}}$ is then manually transferred across the planetary gap, here we nominally set that value to 0.5, which is consistent with the range of values measured from multi-dimensional simulations \citep{Dangelo2002,Lubow2006}. We note the position at which the measurement of the steady-state viscous flux is determined is typically three times the planetary separation; however, sometimes, the disc becomes truncated by photoevaporation inside this radius. Hence here we choose one and a half times the planet's separation. While we could include a parameterised model for how the planet additionally accretes and grows in mass \citep{li2021}, this process is again uncertain. Thus, in order to isolate physical effects in our parameter study, we choose to fix the planet's mass for the duration of the simulations. In our initial calculation, the planet is inserted at 4 AU, with a mass of 1 M$_J$ at the beginning of the simulation. 

\begin{figure*}
    \centering
    \includegraphics[width=\textwidth]{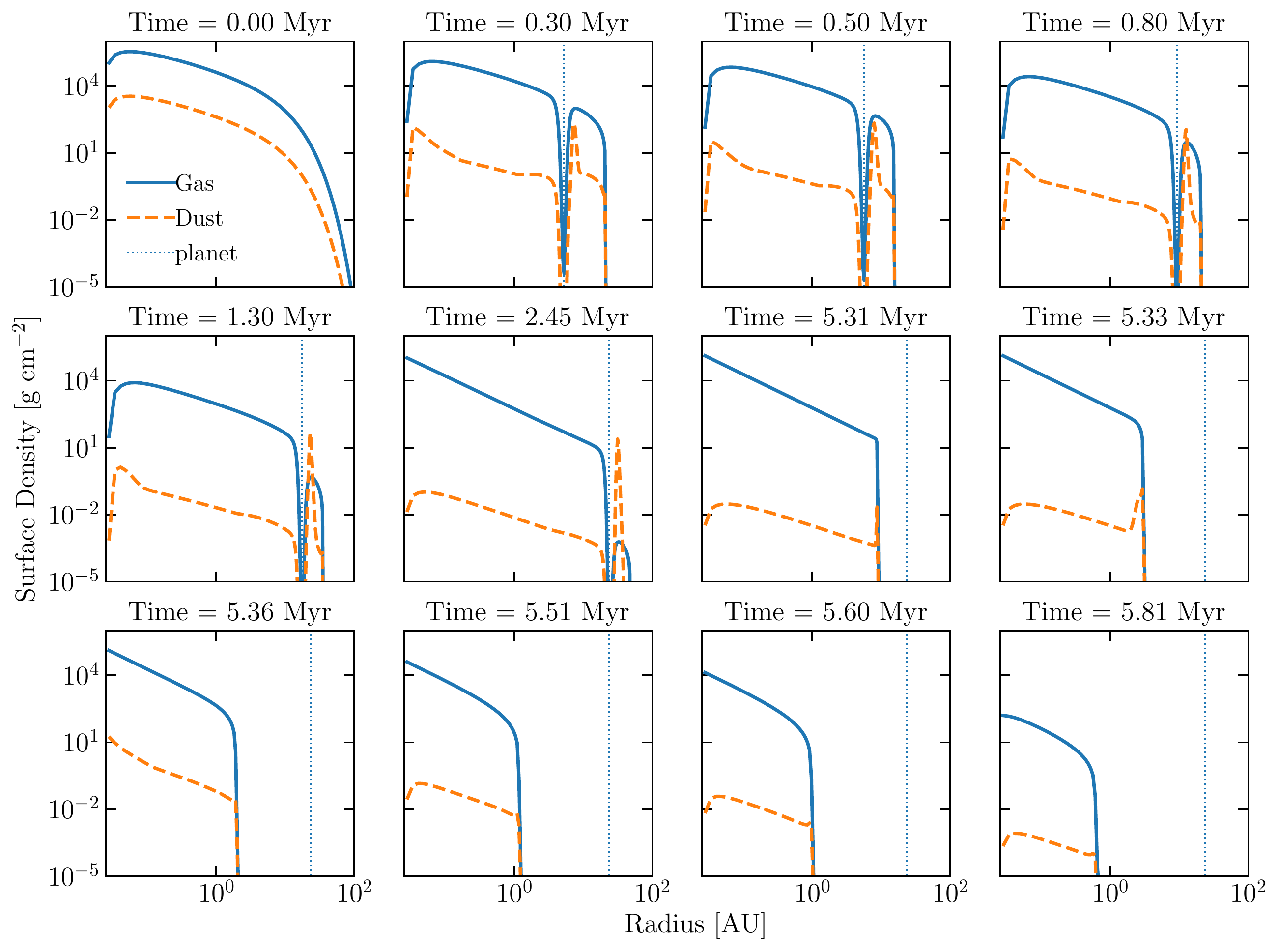}
    \caption{Dust and gas surface density shown at different snapshots in time throughout the disc's evolution. The planet's location is indicated by the position of the vertical dotted line. The dust-to-gas ratio is significantly enhanced in the outer disc by the giant planet, suppressing the photoevaporative mass-loss. Between the snapshot at 1.3~Myr and 2.45~Myr the disc switches from an accretion to a decretion disc (signified by the change in surface density profile in the inner regions). Once the outer disc is depleted of gas (at $\sim 5$~Myr), the dust is assumed to be blown away, and the disc starts rapidly photoevaporating from outside in, lasting a few $10^5$~years before it eventually disperses. }
    \label{fig:evolve}
\end{figure*}

\begin{figure}
    \centering
    \includegraphics[width=\columnwidth]{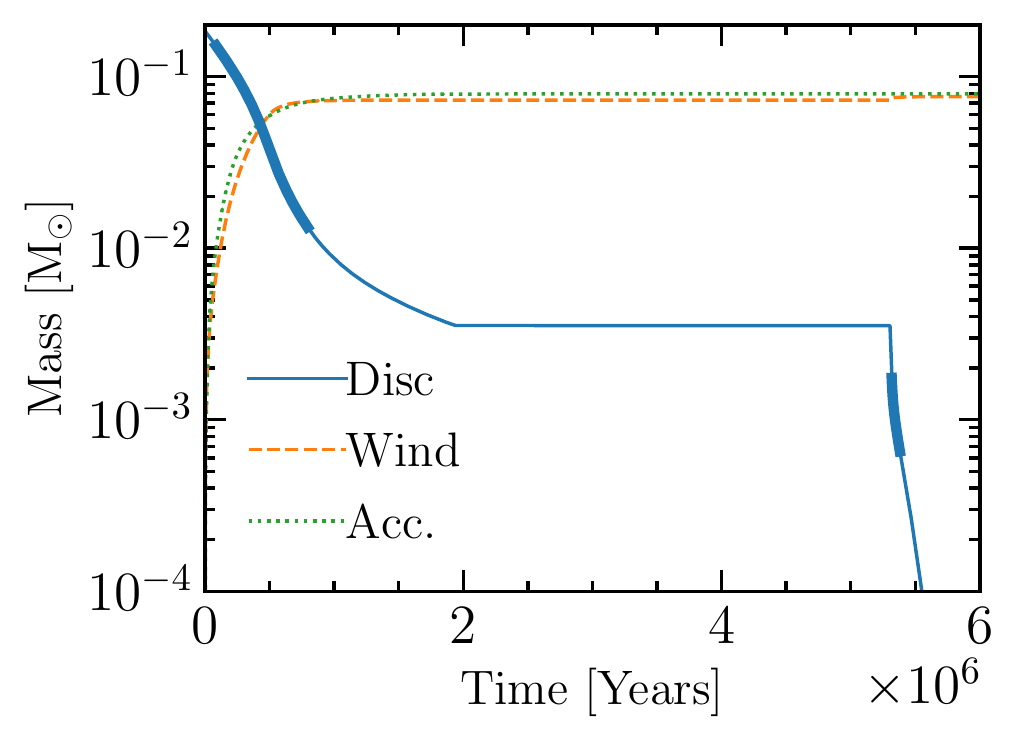}
    \caption{ Disc mass and mass lost to accretion and photoevaporation as a function of time for the evolution calculation shown in Figure~\ref{fig:evolve}. The combination of the planetary and magnetic torque holds the disc at a constant mass from $\sim 2-5$~Myr. When the disc mass line is thick, the photoevaporative mass-loss rates are sufficient to match the required Br-$\gamma$ fluxes of the G-clouds. }
    \label{fig:standard_masses}
\end{figure}

\begin{figure}
    \centering
    \includegraphics[width=\columnwidth]{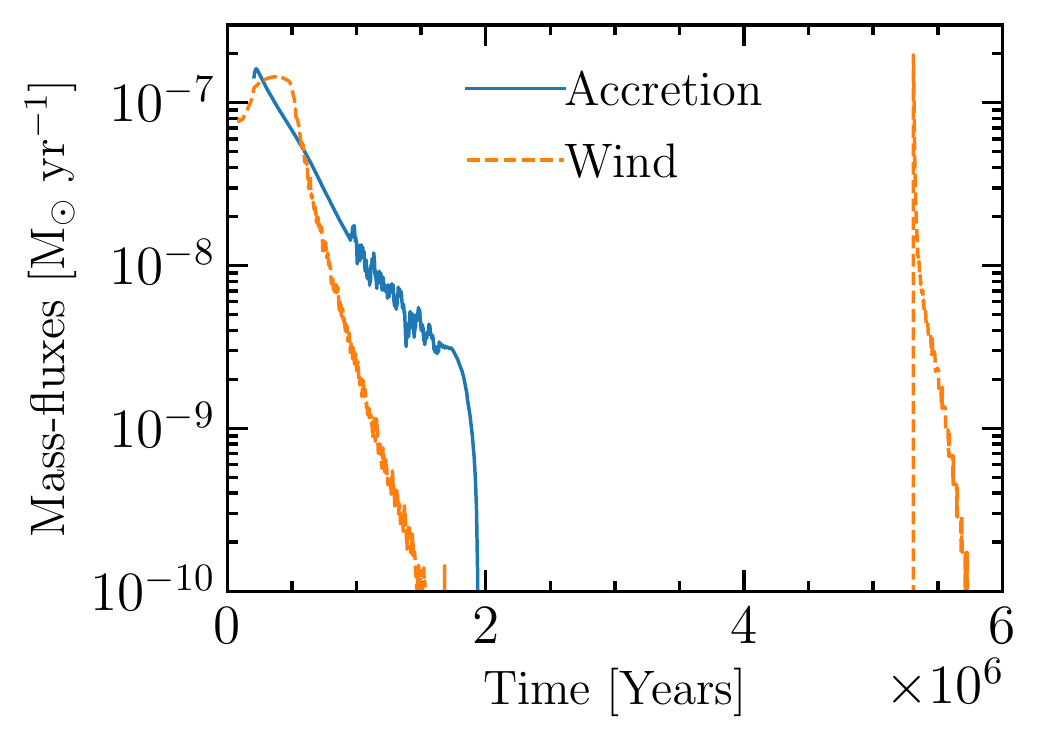}
    \caption{The mass accretion rate (solid) and photoevaporative mass-loss rates (dashed) for the disc evolutionary calculation shown in Figure~\ref{fig:evolve}. The variable accretion period occurs during a disc-locking phase, where the mass accretion rate and stellar spin are modulated to fix the disc's inner edge close to co-rotation.}
    \label{fig:standard_fluxes}
\end{figure}

The evolution of the dust and gas surface density are shown in Figure~\ref{fig:evolve} as snapshots of their profiles at relevant points in the evolution. In addition, we show the evolution of the disc mass and the mass lost to accretion and the wind in Figure~\ref{fig:standard_masses}. Finally, the accretion rate onto the star and the photoevaporation rate is shown as a function of time in Figure~\ref{fig:standard_fluxes}.

These Figures show that dust is trapped outside the planet's orbit in this scenario, locally enhancing the dust-to-gas ratio at about twice the planet's semi-major axis \citep{Pinilla2012}. Photoevaporation will then remove disc material until the outer disc edge approaches the region where the dust-to-gas ratio is enhanced, suppressing the mass-loss rate. While this is occurring, the disc will continue to accrete onto the star, evolving to the point where the stellar magnetic field is strong enough to prevent accretion. As the star is preventing accretion, angular momentum will be transferred to the disc, causing it to expand, which will ultimately cause the planet to migrate outwards. In this way, the evolution of the system will be governed by the viscous time of the disc, which as the planet's orbit expands, will increase. Ultimately, we expected the disc could survive for several Myr in this scenario. However, since we have suppressed or shut off photoevaporation, it will not appear as a G-cloud at this stage. What we find in our simulations is that our manually implemented leakage slowly drains the outer disc of gas. Once the outer disc is fully drained of gas, the dust particles are assumed to be blown out, and the EUV radiation can penetrate to the disc inside the planet's orbit. This disc is now dust-depleted, and photoevaporation can proceed again at an enhanced rate, which can exceed a rate of $10^{-7}$~M$_\odot$~yr$^{-1}$. Photoevaporation now destroys the disc on a timescale of a  $\sim$10$^5$~years with the evolution looking similar to the case of a non-accreting disc discussed in Section~\ref{sec:non_acc} (Figure~\ref{fig:no_acc}). This agreement is not surprising as the initial conditions for our non-accreting discs are similar to the disc properties once the disc material outside the planet has been depleted. We do wish to highlight that this switch from a disc which is not accreting or photoevaporating to one which appears like a G-cloud is slightly pathological in our modelling in that is arising due to the implementation of a leaky planetary gap. If we did not include any leakage across the gap in our simulation, the simulated disc could essentially live for a very long time. We wish to highlight that in our simulation, the leakage prescription moves disc material from the outer disc to the inner disc even in the decreation disc phase, where the net viscous flow is outwards; however, the gas disc profile in the vicinity of the planet promotes inward viscous flow. Thus, it may be that, in this case, disc material does not flow inwards. Nevertheless, this pathological transition from a non-accreting, non-photoevaporating disc to a rapidly photoevaporating G-cloud phase in our simulations is not a particularly concerning problem. In section~\ref{sec:rates} we discuss several other processes that are prevalent in the Galactic Centre that will naturally cause the disc to transition to a G-cloud whether the leakage works as we implemented or not. In this discussion, we actually suggest the more likely scenario to transition into the G-cloud phase is to remove the planet through dynamical encounters. 

\subsection{Sensitivity to parameters}\label{sec:param}

\begin{figure*}
\centering
\includegraphics[width=\textwidth]{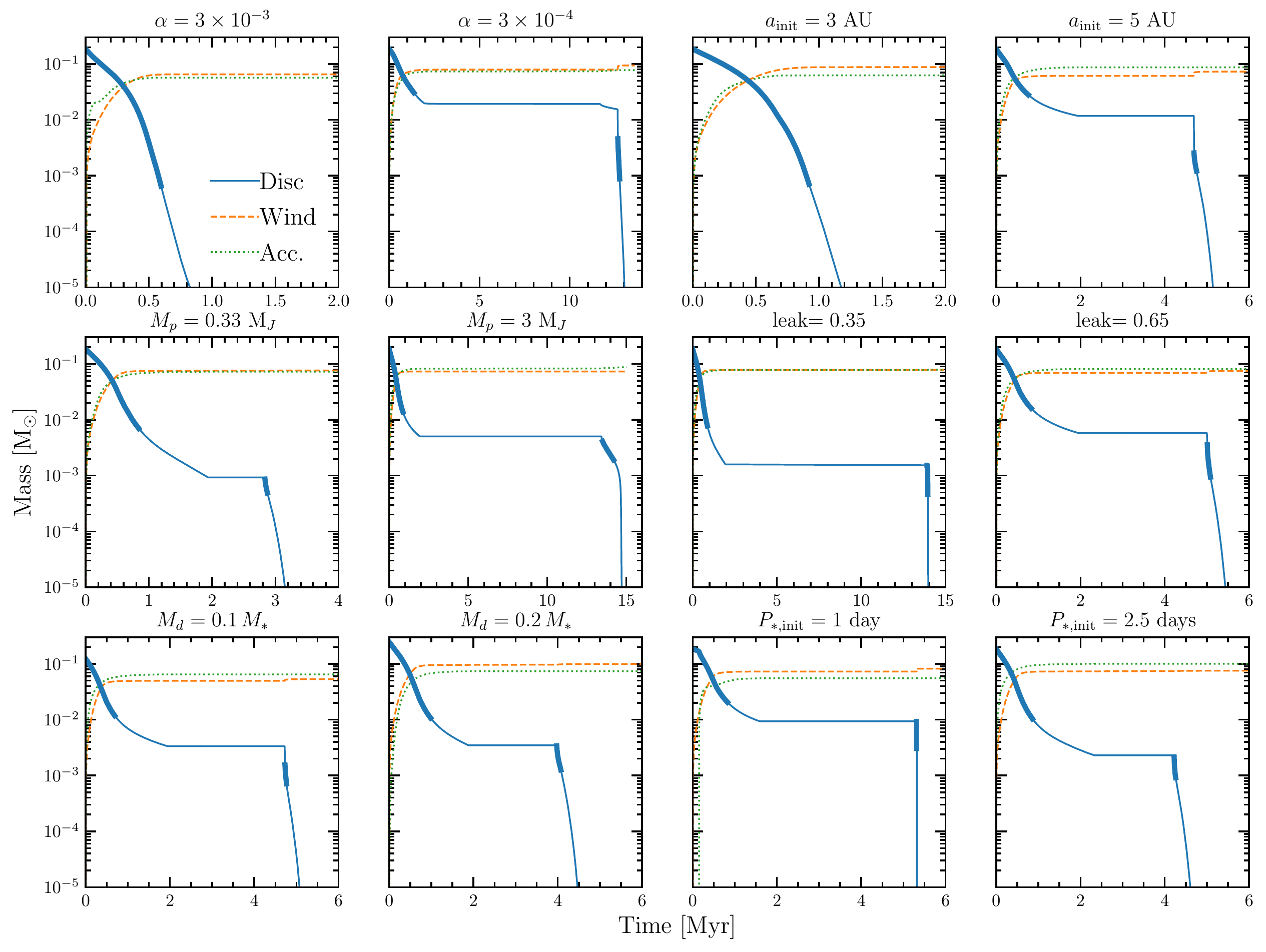}
\caption{This figure shows the evolution of the disc's mass and the cumulative masses accreted onto the star and lost to the photoevaporative wind. Each panel varies a single one of the listed parameters away from or ``nominal'' case shown in Figure~\ref{fig:standard_masses}. The thick portions of the disc mass line indicate the time at which the disc would be photoevaporating sufficiently strongly to appear as a G-cloud in Br-$\gamma$.  }\label{fig:param}
\end{figure*}

Since we have identified a successful pathway to explain the G-clouds originating from circumstellar discs around T-Tauri stars formed in the most recent star formation episode a few Myr ago, it is worth checking how sensitive to the parameters the scenario is. Specifically, we wish to determine whether the scenario is fine-tuned or not. Here we vary the viscous $\alpha$, initial planet mass and separation, the leakage parameter across the gap, the planet's disc mass and the initial spin period of the star. The resulting evolution of the disc mass and the times the disc will appear as a G-cloud are shown in Figure~\ref{fig:param}. We find that, in general, the scenario is not fine-tuned and while the overall disc lifetime and the length of the G-cloud phase (ranging from $\sim 10^5$ to $10^6$~years) are sensitive to the parameters, the resulting evolutionary pathway is often the same. We find, as one would expect, lower viscosities,  higher planet masses, and smaller leakages result in longer disc lifetimes. Additionally, the star's initial spin period and viscosity play a strong role in determining the disc mass at which accretion is shut off, with faster spin rates and lower viscosities shutting off accretion at higher masses. 

Our parameter study does identify a couple of ``failure modes'' of the scenario, which arise from the inability to shut off either accretion and/or photoevaporation. The planet with an initial starting radius of 3~AU shown in Figure~\ref{fig:param} indicates a case where the planet is able to migrate inwards towards the star, becoming a hot Jupiter. In this example, photoevaporation can never shut off and ultimately destroys the disc. Another case is shown for a viscous $\alpha$ value of 3$\times10^{-3}$, here the viscosity is high enough that accretion can continue onto the star until later times. Finally, although not shown, if we make the initial spin period longer than $\sim 4-5$ days, accretion can only be prevented at extremely low disc masses, and the remaining disc reservoir is insufficient to explain the mass-loss rates of observed G-clouds. {\rc These failure modes highlight the need for the disc to become a decretion disc at sufficiently high mass. Otherwise, the planet continues its migration towards the star, and the dust trap migrates inward, allowing accretion and photoevaporation to proceed.} Ultimately, we expect star formation in the Galactic Centre to give us a range of different initial star/disc/planet parameters, many of which evolve into G-clouds at different times, giving a continual supply of G-clouds that last $10^5$ to a few $10^5$ years each over a period of Myrs. 

Additionally, it is worth investigating this evolutionary picture for discs in more typical star-forming environments, as it may play a role in resolving the propyld lifetime problem.

\section{Transition to a G-cloud} \label{sec:rates}
We have shown in the previous section that the presence of a giant planet allows the disc to survive a few Myr in the Galactic Centre. However, by using the giant planet to suppress photoevaporation, allowing the disc to survive, the disc must transition to a G-cloud phase at late times. As a basis for the following discussion, we consider that there are currently around 5-10 observed G-clouds ($N_G$) in the Galactic Centre \citep[e.g.][]{Peissker2020,Ciurlo2020}. 

Additionally, extrapolating the observed IMF from the young cluster in the Galactic Centre \citep{Lu2013}, we estimate roughly 2000 T Tauri stars ($N_{TT}$) with masses between 0.3 and 1.5~M$_\odot$ were born in the young disc. These young stars roughly represent the progenitor population of the G-clouds. This number is in agreement with the value estimated by \citet{Nayakshin2005}, assuming a fraction of the diffuse X-ray emission from the Galactic Centre is produced by T Tauri stars, as demonstrated by \citet{Lu2013}. Certainly, more than a few thousand T Tauri stars are ruled out by the X-ray emission constraints of \citet{Nayakshin2005}.

In the scenario where we use an embedded planet to suppress photoevaporation, adopting  a G-cloud lifetime of $t_G$, we may write the number of G-clouds at any instant as:
\begin{equation}
    N_{\rm clouds} \sim f_p N_{TT} \Gamma_{G,*} t_G
\end{equation}
where $f_p$ is the fraction of the stars that form planets that allow the disc to survive photoevaporation and $\Gamma_{G,*}$ is the rate at which an individual planet-hosting star becomes a G-cloud. Assuming the production of G-clouds has roughly been uniform over the age of the young cluster ($t_{\rm age}$), we have $\Gamma_{G,*} \lesssim 1/t_{\rm age}$.  This implies that the G-cloud lifetime must be $\gtrsim f_p^{-1}\,1.5\times 10^4$~yr for $t_{\rm age}=3$~Myr and $N_{\rm clouds}=10$. Taking $f_p\sim 0.1$, the value obtained from exoplanet surveys, for a fraction of stars that host giant planets on AU scale orbits \citep[e.g.][]{Cumming2008,Winn2015,Fernandes2019} we arrive at a G-cloud lifetime of order $\sim 10^5$ yr. This value is exactly what we find in our calculations. Thus, provided T Tauri stars in the Galactic Centre can form giant planets at a similar rate to those studied in the Solar Neighbourhood then our presented scenario can produce G-clouds at the required rate. 

As discussed previously, in order to appear as a G-cloud the disc must exit the phase where photoevaporation has been shut off by the giant planet's dust trap. Below, we consider three possible scenarios and discuss their relative likelihood. 

\subsection{Draining of the outer disc}
The first scenario is the one identified in our simulations, that the gas in the outer gas disc is destroyed both by draining into the inner disc via accretion and photoevaporation. This scenario is the most difficult to calculate as the gas transfer rate across the gap is poorly known, even in the more standard case of a planet embedded in an extended disc. As discussed in section ~\ref{sec:numerical}, in our simulations, we parameterised this transfer rate in terms of the fraction of the mass accretion rate a steady-state disc would provide with a given surface density. In the majority of our simulations, we found that accretion draining dominates over photoevaporation (as photoevaporation is significantly suppressed at high dust-to-gas ratios {\rc since the dust absorbs the EUV photons}); hence in our estimates below, we neglect photoevaporation.

The disc mass in the outer regions trapped outside the planet is given by approximately:
\begin{equation}
    M_d \sim 2\pi R H_R \Sigma 
\end{equation}
where $H_R$ is the radial size of the gas disc exterior to the planet. The accretion draining rate is given by:
\begin{equation}
    \dot{M} = f_{\dot{M}} 3\pi \nu \Sigma
\end{equation}
hence the depletion timescale is:
\begin{equation}
    t_{\rm dep}\sim \frac{1}{\alpha f_{\dot{M}}} \left(\frac{R}{H}\right)\left(\frac{H_R}{H}\right)\Omega^{-1}
\end{equation}
Our simulations indicate that $H_R/H \sim3-5$ (Figure~\ref{fig:evolve}), thus for $\alpha=10^{-3}$ if $f_{\dot{M}}$ is of order a few tenths then the depletion timescale is a few Myr at around 10 AU. 

Thus, if $f_{\dot{M}}$ is of order a few-tenths then it will deplete the outer discs on the required timescale of a few~Myr. If $f_{\dot{M}}$ is $\sim 1$ or larger, then the discs will deplete too rapidly and if $f_{\dot{M}} \ll 1$ then discs will not deplete quickly enough. It is unclear if this is a fine-tuning problem, since $f_{\dot{M}}$ of a few tenths is roughly in agreement with values provided from simulations of planet-disc interactions (albeit in more extended discs, \citealt{Lubow2006}).

\subsection{Tidal stripping of the planet by Sgr A*}

Since our T Tauri scenario results in an outwardly migrating planet, if the planet were to migrate to sufficiently large distances it would be tidally stripped by the black hole, allowing the transition to a G-cloud phase.  

The disc interior to the planet is not allowed to accrete, and hence will behave like a steadily expanding decretion disc \citep[e.g.][]{Pringle1991,Nixon2021} with a constant outward angular momentum flux of $F_J = 3\pi \nu \Sigma h$. This angular momentum flux will be absorbed by the planet causing it to migrate outwards at a rate:
\begin{equation}
    \frac{a_p}{\dot{a}_p} \sim \frac{M_p}{M_{\rm d,in}} t_\nu
\end{equation}
We find the tidal stripping scenario to be the least likely scenario for creating the G-clouds. At a distance of $0.1$~pc from Sgr A* the Hill radius for a solar-mass star is approximately 80~AU, meaning the planet will be tidally stripped at a distance of roughly 40~AU. The viscous time-scale at $\sim 40$~AU is long ($t_{\nu}\sim 1/\alpha (R/H)^2 \Omega^{-1}$), necessitating a low-mass planet. However, the planet's mass cannot be arbitrarily low. At too low planetary masses the perturbed disc structure is unable to trap dust exterior to its orbit. This happens at the ``pebble-isolation mass'' \citep{Lambrechts2014}, which is closely related to the planetary thermal mass. For a typical disc around a solar mass T Tauri star, the disc's aspect ratio is approximately $H/R\sim 0.07$ at 40 AU. For $\alpha=10^{-3}$, \citet{Bitsch2018} find the pebble-isolation mass to be $\sim 0.2$~M$_J$. Our simulations indicate that accretion is shut off and the planet begins its outward migration for disc masses of few Jupiter masses, indicating a minimum planetary migration timescale of $\sim 1~$Myr. This timescale would be just about compatible with the observations. However, it is unlikely the planet could remain at this low mass for the entire evolution. sub-Jovian mass planets are subject to run-away accretion in a gas-rich environment \citep[e.g.][]{Pollack1995} and can accrete rapidly reaching Jovian masses, slowing the outward migration.  In addition, the photoevaporation rates for a 40~AU disc are extremely large ($\sim 10^{-6}$~M$_\odot$~yr$^{-1}$, resulting in a lower disc mass once it is rapidly photoevaporated down to radii of a few AU and shorter disc lifetimes.  Thus, while roughly compatible with the required observational rates, tidal stripping of the planet by the black hole might be fine-tuned, unless the progenitor stars have pericentre distances less than 0.1~pc. Finally, we note that by the time planet has migrated out to $\sim 40$~AU, the disc/planet system might have extracted so much angular momentum from the star that it is no longer spinning rapidly enough to prevent accretion. Ultimately, this may cause the disc to begin accreting and potentially allow the planet to migrate inwards again. 

\subsection{Removal of the planet via encounters}

The Galactic Centre is a reasonably dense stellar environment, with around a stellar mass of $10^6$~M$_\odot$ inside 1~pc \citep[e.g.][]{Genzel2010}. Therefore, the encounter rate with another object is high. The majority of these objects are old main sequence stars and compact remnants that have migrated towards the Galactic Centre \citep[e.g.][]{Miralda2000,AlexanderT2009,Linial2022}. An encounter with a closest approach distance similar to the orbital separation of the planet is likely to be highly disruptive. Such close encounters often unbind the planetary companion from the host star and tidally truncate the disc \citep[e.g.][]{Clarke1993,Pfalzner2005,Malmberg2007,Davies2014}. Due to the sheer number of old stars and remnants, they are likely to dominate the close encounters, rather than other young stars\footnote{$H/R$ of the young disc would need to be $\sim 10^{-3}$ for the local density of young stars to be similar to old stars.}. This population of old stars and remnants is relaxed, and thus has a velocity dispersion ($\sigma_d$) of order the Keplerian speed around the black hole, significantly higher than the escape velocity of the planet. As such, the collision cross-section on AU scales or greater is simply the geometric cross-section (the Safronov number $\ll 1$). In this case, the encounter rate is $\Gamma_{\rm en}\sim n_{\rm old}\pi b^2 \sigma_d$, where $n_{\rm old}$ is the density of old stars/remnants and $b$ is the impact parameter of the encounter. \citet{MiraldaEscude2012} provide estimates of the density of old stars and stellar mass black holes in the Galactic Centre, since their velocity dispersion are similar and the density of old stars exceeds that of the remnants, close encounters on AU scales are dominated by those with the old stars. \citet{MiraldaEscude2012} argued that the relaxed old stars will approach a $r^{-3/2}$ density profile (in agreement with the calculations of \citealt{Vasiliev2017}), thus the encounter rate for an individual T Tauri star is:
\begin{equation}
    \Gamma_{\rm en} \sim 1~{\rm Myr}^{-1} \left(\frac{r}{0.1~{\rm pc}}\right)^{-2} \left(\frac{b}{5~{\rm AU}}\right)^2\label{eqn:encounter_rate}
\end{equation}
where we have followed \citet{MiraldaEscude2012} in normalising the old star density profile assuming that the population consists entirely of 1~M$_\odot$ stars. Now since an order unity fraction of close encounters with impact parameter $b$ can strip a planet with an orbital semi-major axis of order $b$, then the derived encounter rate is sufficient to remove planets on 1-10~AU orbits on Myr timescales, transitioning the planet-disc system into a G-cloud. The $b^2$ scaling necessitates that encounters that completely destroy the disc, resulting in too little mass to form a G-cloud are rare. Alternatively, instead of a single disruptive encounter, repeated longer-distance interactions with other objects could ``evaporate'' the star-planet system. Frequent interactions cause the planet's semi-major axis to grow by giving energy to the planet, eventually unbinding it. \citet{Rose2020}, estimate this evaporation timescale to be $\sim 10^6$~years for objects with distance from Sgr A* to be of order 0.1pc.   Hence, based on the discussion above, we suspect that either a single disruptive, close encounter with a nearby old star or frequent interactions with other stars in the Galactic Centre are the most likely mechanisms for transitioning the disc into a G-cloud.


\section{Pressure confinement: the special case of G2}\label{sec:pressure}

The G2 cloud deserves special attention. G2 is the most extreme of the identified G-clouds with an eccentricity of $\sim 0.98$ \citep[e.g.][]{Gillessen2012,Plewa2017} resulting in a pericentre distance that takes it to within 150~AU of the black hole. This extremely close passage has two important consequences for our scenario. In our scenario, the G-clouds have lifetimes longer than their Keplerian orbital periods, which range between 100 and 1000's years \citep[e.g.][]{Gillessen2012,Ciurlo2020}, thus, it is likely G2 has performed many pericentre passages already. The Hill radius for a Solar mass star at a distance of $\sim 150$~AU around a $4\times 10^6~$M$_\odot$ black hole is only around 1~AU. Repeated pericentre passages are likely to result in tidal truncation of the disc down to a similar size scale, with the exact tidal stripping radius depending on the orientation of the disc's angular momentum axis relative to its orbital angular momentum axis around Sgr A*. Figure~\ref{fig:photo_rates} indicates that a disc with a dust-to-gas ratio of order $\sim 10^{-4}$ can explain the Br-$\gamma$ flux of G2 with a disc radius of $\sim 1.5$~AU, slightly larger than the nominal 1~AU truncation radii. We speculate that this mild tension is resolved by the fact the pericentre passage of G2 occurs in a fairly high-pressure environment, and as discussed in Section~\ref{sec:brgamma}, our Br-$\gamma$ luminosities are underestimates for pressure-confined outflows. The vicinity of Sgr A* contains dense gas, with densities of $\gtrsim 10^4$~cm$^{-3}$ expected in the vicinity of pericentre passage \citep[e.g.][]{Xu2006, Murchikova2019}. While these densities are too low to result in significant ram-pressure stripping over the G-cloud's lifetime, they may provide additional pressure support during pericentre passage preventing the disc from being tidally truncated to small sizes. High-resolution simulations of the compact-source scenario for G2, have a minimum resolution of $\sim 5~$AU, thus new, higher resolution simulations are required to fully explore the truncation of a disc at pericentre passage. 

\begin{figure}
    \centering
    \includegraphics[width=0.92\columnwidth]{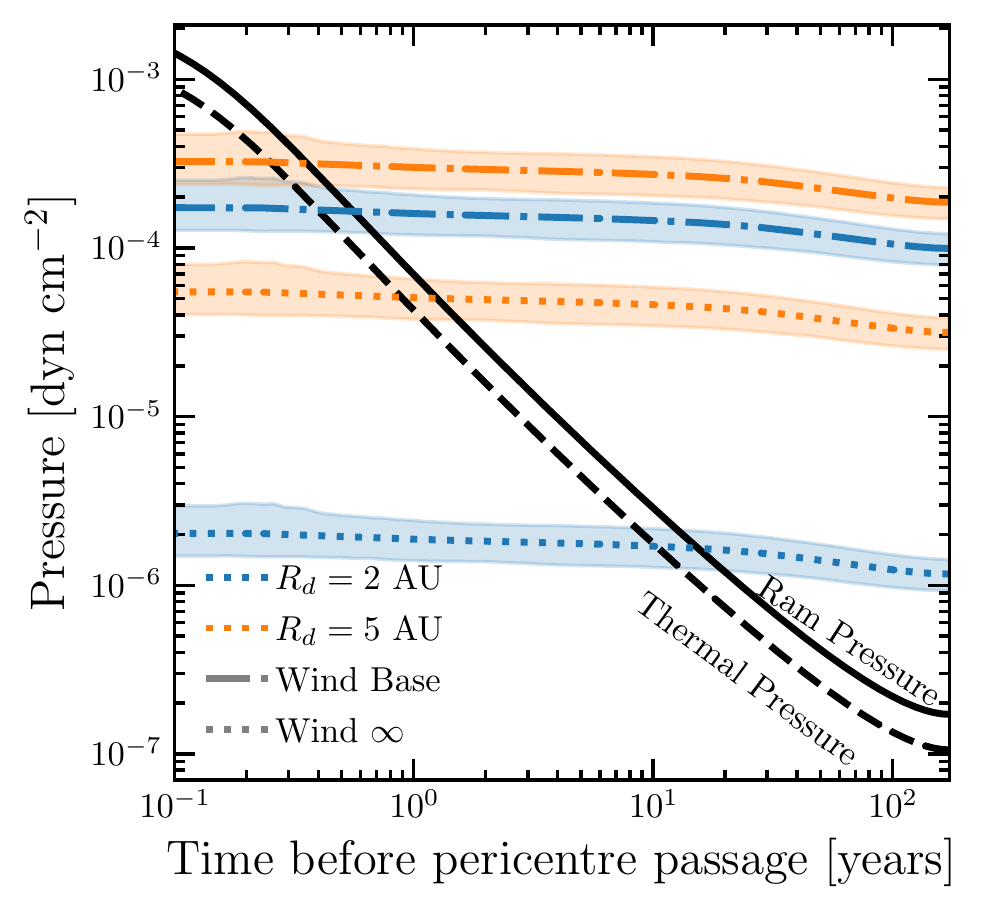}
    \caption{The solid and dashed lines show the ram pressure (assuming a static medium) and thermal pressure acting on G2 as a function of time before pericentre passage. The dot-dashed line shows the thermal pressure at the photoevaporative wind base and the dotted line shows the ambient pressure to prevent the wind from escaping for disc sizes of 2 and 5 AU. The shaded regions show those arising from the 5--95\% EUV fluxes shown in Figure~\ref{fig:EUV_flux}. We use fluxes for objects perpendicular to the O/WR disc; choosing the parallel model does not change the times at which the wind becomes pressure confined. G2's spends most of its time able to drive a hydrodynamic photoevaporative wind; however, on approach to pericentre the wind region becomes pressure confined, and finally, at pericentre passage, the disc becomes pressure confined.  }
    \label{fig:pressures}
\end{figure}

This realisation that the ambient pressure environment in the Galactic Centre matters raises an important question: can you actually drive a photoevaporative outflow? To obtain a photoevaporative outflow, the external pressure at large radii needs to be sufficiently low that the outflow can pass through a sonic point \citep[e.g.][]{Parker1958,Lamers1999}. If the external pressure is too large, the photoionized gas will be held in hydrostatic equilibrium by the external environment. The external pressure will be too weak to prevent a hydrodynamic outflow if the hydrostatic pressure of the photoionized gas from the disc exceeds the external pressure at large radii, or:
\begin{equation}
    P_{\rm ext}< P_{w,\infty} \approx P_b\exp\left(-\frac{GM_*}{c_s^2R_{\rm disc}}\right)
\end{equation}
where $P_b$ is the pressure at the base of the photoionized region at the edge of the disc. In Figure~\ref{fig:pressures}, 
we show the external thermal and ram pressures (assuming that the external medium's relative velocity with G2 is just the Keplerian value of G2 orbit), adopting the orbital parameters of G2 from \citet{Plewa2017}. Additionally, we use our photoevaporation models and EUV flux estimates to show both the hydrostatic pressure of the photoionized region at large distances ($P_{w,\infty}$) and the pressure of the base of the photoionized wind. What this comparison demonstrates is that for most of its life, the external pressure is too weak to prevent a hydrodynamic outflow. However, once G2 begins its plunge towards Sgr A*, the photoevaporative outflow is pressure confined a few to 10s years before pericentre passage, with smaller discs being confined earlier. At pericentre passage, the entire photoionized region has a pressure lower than the external environment. Hence, at pericentre passage, it is likely that the photoionized gas around the disc is ram pressure stripped. It would then take $\sim R_g/c_s\approx 4~$ years for the photoionized region to be reestablished around the disc. Thus, we hypothesize the following orbital evolution for G2 in the photoevaporative disc scenario. At apocentre, the disc behaves like a standard photoevaporating disc; however, as it falls towards Sgr A* the outflow would be shutoff and would result in a pressure confined gas cloud of size $\sim 100$~AU around the disc/star, around 10~years before pericentre passage for a disc of a couple of AU in size. As G2 approaches pericentre, this gas cloud would be further pressure confined and tidally stripped. Since the change in external pressure is fast compared to the sound crossing time of the photoionized gas cloud it is likely to highly dynamic. At pericentre passage, the photoionized region is stripped, allowing a new photoionized region to be created on a timescale of a few years; once G2 is sufficiently far from Sgr A* (after $\sim 10$~years), the photoevaporative outflow will no longer be pressure confined and wind will be re-established. Within $\sim 100$~years before the next pericentre passage, the wind will output a few Earth-masses of ionized gas, sufficient to explain the Br-$\gamma$ fluxes every orbit (as if the photoionized gas is stripped every pericentre passage it must be replenished every orbit). We additionally add the pressure confinement helps ease the tension between the disc's size required to match the Br-$\gamma$ flux and truncation radius.  Being pressure confined region results in a higher density at large radii than an expanding outflow, and as such will give higher Br-$\gamma$ fluxes for the same disc radius. 

This evolutionary sketch perhaps explains a couple of the mysteries of G2. The pressure-confined gas cloud scenario (without a central compact source) nicely explains the position-velocity structure of the Br-$\gamma$ emission as G2 approached pericentre \citep[e.g.][]{Schartmann2015}. However, since the object contains a central source that can reform a pressure-confined photoionized cloud after pericentre passage, one can explain how a G2 survived pericentre passage \citep[e.g.][]{Plewa2017}. Thus a photoevaporating disc behaves like the pressure-confined diffuse gas cloud scenario in the decade preceding pericentre passage, but like the compact source scenario during pericentre passage. As mentioned above, the sound crossing time for the photoionized region is longer than the timescale on which these transitions are likely to happen. Thus, a 3D simulation must be performed to explore the dynamical evolution of a disc and its photoionized region as it orbits Sgr A*.

Further, it has been suggested that G2 is potentially slowing in its orbit as if it is experiencing a drag from the surroundings \citep[e.g.][]{Gillessen2019}. This is problematic for the compact source scenario as it is difficult to understand how an object with a mass of $\sim 1$~M$_\odot$ would experience sufficient drag to slow it down enough for a deviation from the Keplerian orbit to be observed. However, as described above, we may expect the photoevaporative wind to be stripped from the star during pericentre passage, and it takes a few tens of years for the disc to relaunch a new wind. The photoionized region itself has a mass of only a few Earth masses, sufficiently low that it could experience a strong deceleration from drag. Thus, we speculate that a solution to this potential paradox is that the G2's pressure-confined photoevaporative region experiences sufficient drag during pericentre passage to either cause it to trail behind the star or even detach entirely, making it appear as if G2 has slowed down. 

Finally, taking the orbital parameters from \citet{Ciurlo2020}, the majority of the remaining G-clouds do not appear to be orbits that take them sufficiently close to Sgr A* for them to become pressure confined. Thus, we suspect they will behave like photoevaporating discs for the entirety of their lifetimes.

\section{Discussion}
In this work, we have studied the photoevaporating circumstellar disc hypothesis for the G-clouds in the Galactic Centre. Through the use of explicit calculation of these discs' secular evolution, we have shown that accretion discs around T Tauri stars that formed in the same star formation epoch that created the young, massive stars cannot survive sufficiently long enough to be observed as G clouds. This brief lifespan is primarily driven by the fact that photoevaporation truncates the discs to small radii (several AU), causing them to accrete rapidly onto their host stars. This process is well understood in even more benign star-forming environments, where photoevaporation brings about more rapid disc evolution, not due to the explicit removal of mass, but the shrinking of the disc's size causing it to evolve faster \citep[e.g.][]{Clarke2007}. Hence, our evolutionary calculations demonstrate that the G clouds are unlikely to be circumstellar discs around young T Tauri stars that follow the typical evolutionary pathway identified in nearby star-forming environments. Since the photoevaporating disc scenario for the G-clouds has many desirable characteristics when compared to observations \citep[e.g.][]{Murray-Clay2012,MiraldaEscude2012}, we instead propose that the G-clouds are not photoevaporating accretion discs, but rather photoevaporating decretion discs, where accretion is prevented by a large magnetic torque from the central star. 
\subsection{The G-clouds as photoevaporating decretion discs around young stars}
Young T Tauri stars are well known to host strong magnetic fields. In our G-cloud scenario where the decretion disc is created by a young star, we have speculated that the removal of disc mass by photoevaporation results in the young star spinning rapidly at an age of a few Myr. This is because as the star contracts towards the main sequence, it accretes less material than in a typical star-forming environment, allowing it to approach breakup velocity. This means that the star is able to magnetically disrupt the disc outside co-rotation while the disc's mass is still significant ($\gtrsim 10^{-3}$M$_*$). The magnetic torque prevents accretion and results in the disc expanding as a decretion disc \citep[e.g.][]{Pringle1992}, where both angular momentum {\it and} mass are transported outwards. Without any additional processes, this outward transport of mass causes the disc to undergo more vigorous photoevaporation resulting in disc lifetimes that are too short to explain the G-clouds. However, a forming giant planet, which can trap dust exterior to its orbit, significantly suppresses photoevaporation once the dust-to-gas mass ratios exceed $\sim 0.1$ exterior to the planet. The suppression of photoevaporation occurs because dust particles dominate the absorption of EUV photons, resulting in low densities of photoionized gas and weak mass-loss rates. At this stage, these non-accreting, non-photoevaporating discs can survive for $\gtrsim 5-10$~Myr as the planet migrates outwards slowly.

Ultimately, we suspect encounters with old stars in the Galactic Centre will eject or unbind the planet, removing the dust trap, and allowing the disc to begin photoevaporating again. It is this photoevaporating remnant we hypothesise that creates the G-cloud signature, and they have lifetimes of $\sim 10^5$~yrs. 
\subsubsection{Planet formation in Galactic Centre}

In order for the T Tauri star scenario to work, we need a planet to form in the disc that is sufficiently massive to trap dust exterior to its orbit ($\gtrsim 0.2$~M$_J$). While the required fraction of discs that form such planets is consistent with both the giant planet occurrence rate on AU scales and the fraction of protoplanetary discs that appear as ``transition'' discs in nearby star-forming regions, it's unclear whether planet formation in the Galactic Centre will proceed in a similar manner. The tidal effects of the black hole or the impulses injected from more frequent close encounters may increase the velocity dispersion in any planetesimals, reducing the efficiency of planetary core formation. Alternatively, the smaller disc radii would result in more rapid dust growth at the disc's outer edge, giving rise to a higher pebble flux (provided significant dust mass is not lost to photoevaporation early -- \citealt{Haworth2018b}), resulting in more efficient planetary core formation. 

Additionally, in the early stages of the disc's lifetime, while it's still self-gravitating, planet formation via direct collapse could occur. A disc that is stable to fragmentation and planet formation, can be made to undergo fragmentation and planet formation by a tidal perturbation \citep[e.g.][]{Pfalzner2007,Meru2015, Cadman2022}. This mechanism is particularly relevant for our scenario, as it causes fragmentation in the inner regions of the disc \citep{Meru2015}, allowing giant planet formation to occur close-in, early in the disc's lifetime. We speculate that both, the tidal forcing from the black hole and stellar flybys on 50~AU scales which occur frequently (Equation~\ref{eqn:encounter_rate}), could result in enhanced giant planet formation through gravitational fragmentation. Ultimately, planet formation simulations that account for the unique properties of the environment need to be performed before one can assess whether giant planet formation is more or less likely in the Galactic Centre.

\subsection{The merger scenario}
The stellar merger scenario has been suggested as a possible origin for the G-clouds \citep{Witzel2014,Prodan2015,Ciurlo2020}, where the dense Galactic Centre environment causes binaries to merge frequently. Most of these works focus on calculating the properties and frequency of such mergers, what is not clear is how exactly such a merger product creates a G-cloud signature. It could either be the merger product itself or, as we explore here, a photoevaporating decretion disc. Thus, our discussion and rate calculations are focused on exclusively on disc origin.  Provided the merger remnant created is rapidly spinning and magnetised, then any bound high-angular momentum material left over in the collision is likely to form a disc. Like the T Tauri scenario, the strong magnetic torque prevents the disc from accreting, causing it to expand and photoevaporate. We show that such a disc could live for a few $10^5$~years, potentially explaining the G-cloud populations. If the progenitor binaries formed in the most recent star-forming episode, we estimate roughly $\sim 100$ potential binaries formed, a sufficient population to explain the G-clouds if they can all be made to merge in a few Myrs. While current calculations do not indicate such high merger rates \citep[e.g.][]{Stephan2016}, there are other dynamical processes that could be included to enhance the merger rate. 
\subsection{Limitations and further work}
The goal of this work has been to explore the photoevaporating circumstellar disc scenario as the origin of the G-clouds. In order to span such a large range of possible different individual pathways, we have necessarily simplified our approach. We have focused on simple 1D models of both the photoevaporative flow structure and the secular evolution of the discs. {\rc We did not consider the role of radiation pressure on dust in the photoevaporative outflows. Using the bolometric fluxes expected in the Galactic Centre (Section~\ref{sec:NIR}) the typical radiation acceleration (for $Q\sim 1$) on a 0.1$\mu$m size dust grain is 0.1 cm~s$^{-2}$. This is comparable to the gravitational acceleration from a Solar-mass star on AU scales. However, in the case of an approximately isotropic radiation field (as applicable in the Galactic Centre) this radiative acceleration will be significantly reduced. In addition, the stopping time of the dust is $\sim 10^4-10^5$~s at the sonic point, significantly shorter than the flow timescale ($\sim 10^7-10^8$~s) meaning the dust will be tightly coupled to the gas, and the net radiative acceleration reduced by the dust-to-gas ratio. Thus, we do not expect radiation pressure to be able to affect photoevaporation at low dust-to-gas ratios, although it may affect the dust size distribution. At high-dust-to-gas ratios it may help shut down photoevaporation; however, since photoevaporation drops off at high dust-to-gas ratios due to the absorption of EUV photons, neglecting radiation pressure is unlikely to affect our evolutionary calculations.} No photoevaporation models applicable for compact discs in the high EUV environment of the Galactic Centre had been calculated; since we added the complication of self-consistently including entrained dust into these photoevaporation calculations we took a simplified approach. We considered the photoevaporative flow structure to consist of two isothermal winds, an FUV-heated region that transitions through an ionization front into a EUV-heated outflow. While the isothermal approximation for the EUV heated outflow is probably acceptable, the FUV heated region can have a complicated thermal structure \citep{Adams2004,Haworth2018} with feedback between the entrained dust and outflow structure \citep{Facchini2016}. While the size of the FUV heated region only provides an order unity correction to the mass-loss rates \citep{Johnstone1998}; future, work should consider more realistic temperature and flow structures. This would allow both more accurate mass-loss rates to be obtained and more detailed observational signatures to be computed (such as the metal lines). Additionally, we have assumed the outflows are exclusively photoevaporative. Simulations of discs in the propeller regime \citep[e.g.][]{Romanova2004,Ustyugova2006} have additionally shown that material could be ejected from the inner disc in the form of a wind. Several authors \citep[e.g.][]{Scoville2013,Ballone2018} have considered the G-cloud signature to arise from winds from a young star. Future modelling should include both this potential wind component in addition to the photoevaporative wind. 

One weakness inherent in any secular evolutionary calculation of externally photoevaporating discs, such as those performed in this work, is how to match the outflow solution onto the standard thin-disc evolutionary model, and how to calculate dust entrainment \citep[see discussion in,][]{Sellek2020}. Our approach to dust entrainment was to estimate the dust-to-gas ratio at the outer edge by measuring it in a few cells close to the point where photoevaporation truncates the disc providing a reasonable first approach. Recently, \citet{Owen2021} used a ``slim-disc'' model to show that you can self-consistently compute the flow structure from the disc into the outflow and any size-dependent entrainment of dust grains. However, these calculations were performed in steady-state, and no evolutionary method for the slim-disc approach is available. Since the entrainment of dust grains is critical to the evolution of the disc calculations presented here, this should provide such motivation to develop such an evolutionary model. Additionally, we neglected the dynamical feedback of the dust on the gas which can become important when the dust-to-gas ratio exceeds unity (provided the dust's stopping time is sufficiently short). This leads to localised regions in our disc calculations where the dust-to-gas ratio reaches large values in the centre of the pressure trap. While this is probably not physical for the extremes found in the calculation it is unlikely to strongly affect our results. At high dust-to-gas ratios, the advection of dust into the centre of the dust trap will expel gas (by momentum conservation) weakening the strength of the trap and resulting in a more uniform dust-to-gas mixture in the disc exterior to the planet. This process will reduce the dust-to-gas ratio in the centre of the dust-trap, but actually, enhance the dust-to-gas ratio at the edges of the pressure trap. This is the region where the photevaporative wind is entraining dust particles; thus we suspect the inclusion of dust feedback on the gas would actually decrease the photoevaporation rates from the outer edge when the dust trap is present.    

Finally, the T Tauri star and planet scenario may have reconciled the difference between the pressure-confined diffuse cloud scenario and compact source scenario for G2, by demonstrating that even though a photoevaporating disc is a compact source scenario, while G2 has been observed over the last decade its likely behaved in a manner similar to the pressure confined diffuse cloud scenario. This is because the external pressure within the central 0.01~pc is sufficiently high to pressure confine a photoevaporative wind. However, we have still presented no scenario as to how the progenitor disc of G2 ended up on such a highly eccentric orbit. As argued by \citet{MiraldaEscude2012} single scattering events are likely to be highly distributive and could destroy the disc, and it's unlikely we've just happened to catch G2 on its first plunge toward Sgr A*. One way to increase the eccentricity of G2, without a disruptive single event would be through a secular process; for example through secular resonance sweeping \citep[e.g.][]{Zheng2020}. This possibility should be explored in further work.

\section{Summary}

We have studied the hypothesis that the Galactic Centre G-clouds are photoevaporating circumstellar discs. To calculate the photoevaporation rates, and subsequent emission properties, such as the Br-$\gamma$ flux, we have computed a new suite of external photoevaporation models in the EUV-driven regime. These new models account for the sub-sonic launching of the outflow and account for the entrainment of small dust particles. These models indicate shut-down in photoevaporation at dust-to-gas mass ratios that exceed $\sim 0.1$. This drop in the photoevaporation rates occurs because the dust particles themselves absorb the incident EUV photons, reducing the density of photoionized gas. These photoevaporation flow structures indicate that mass-loss rates of order $10^{-8}-10^{-7}$~M$_\odot$~yr$^{-1}$ are required to explain the observational characteristics of the G-clouds. 

By coupling these new photoevaporation models to numerical evolutionary calculations of both the gas and dust components of circumstellar discs, we have explored various hypothesises for the origin of G-clouds. An accreting circumstellar disc around a young low-mass star formed a few Myr's ago in the most recent star formation episode in the Galactic Centre is ruled out. This arises, not because photoevaporation is able to entirely strip the disc, but rather because photoevaporation truncates the edge causing it to accrete rapidly onto the central star on a timescale of $\lesssim 1~$Myr. 

We demonstrate that is unlikely that the G-clouds are any form of accreting, photoevaporating circumstellar disc, including those formed recently through a merger. This is because of photoevaporative truncation of the disc, which promotes rapid accretion and destruction of the disc. In both the young star and merger hypothesis there is unlikely to be a large enough source population of progenitors to explain the fact we currently observe 5-10 G-clouds. {\rc Instead, if G-clouds are circumstellar discs a large magnetic torque from the central star is required to prevent accretion, causing the disc to evolve as a decretion disc transporting material outwards into the photoevaporating wind. 

If the G-cloud progenitors are young T Tauri stars, to overcome the lifetime issue, photoevaporation must be suppressed. If the disc around a T Tauri star formed a giant planet, which would trap dust exterior to orbit, photoevaporation and accretion could both be suppressed for Myr timescales while the planet slowly migrates outwards. Once the planet is ejected by encounters, which occur on Myr timescales per disc, then the star/disc system would enter a photoevaporating decretion disc that would last up to a few $10^5$~years, while having observational signatures reminiscent of the G-clouds. If T Tauri stars form giant planets at a similar rate to the stars in the Solar Neighbourhood, then this scenario can match the occurrence rate of G-clouds we observe in the Galactic Centre today.  Finally, we demonstrate that if G2 can be explained as a photoevaporating disc, it is likely that during the last few decades of observations, on its approach to Sgr A* its photoevaporating outflow was pressure confined. In this scenario, G2 survival of pericentre passage is explained by the compact-source scenario, while its observational characteristics would match the pressure-confined diffuse cloud scenario. }

\section*{Acknowledgements}
We thank the anonymous referee for comments which improved the manuscript. We are grateful to Andreas Burkert and Smadar Naoz for interesting discussions. JEO is supported by a Royal Society University Research Fellowship. This work benefited from the 2022 Exoplanet Summer Program in the Other Worlds Laboratory (OWL) at the University of California, Santa Cruz, a program funded by the Heising-Simons Foundation. For the purpose of open access, the authors have applied a Creative Commons Attribution (CC-BY) licence to any Author Accepted Manuscript version arising.

\section*{Data Availability}
The disc evolution code used in this work is freely available at \url{https://bitbucket.org/jo276/viscous_panda/}. The simulation results themselves will be shared on reasonable request to the corresponding author.



\bibliographystyle{mnras}
\bibliography{example} 




\appendix
\section{Critical EUV flux for photoevaporation to be EUV controlled}
{\rc We can estimate the critical EUV flux for the transition from an FUV to an EUV driven outflows to occur as follows\footnote{A similar argument was presented by \citet{Johnstone1998}; however it applied to large discs ($R_d\gtrsim 100$~AU) rather than the smaller discs we are concerned with here}: the critical EUV flux is when the ionization front occurs at the sonic point of the FUV heated flow ($\sim GM_*/c^2_{s,FUV}$). Ionization-recombination balance at the ionization front yields a density downstream of the ionization front of $n_{II}=c_{s,FUV}\sqrt{3\phi/G\alpha_bM_*}$, assuming the EUV heated region's density profile falls off as $1/r^2$. Momentum balance across the ionization front (see discussion around Equation~\ref{eqn:momentum}) implies the density upstream of the ionization front is approximately (dropping order unity terms):
\begin{equation}
    n_{\rm I}\sim \frac{c^2_{s,EUV}}{c_{s,FUV}}\sqrt{\frac{\phi}{G\alpha_bM_*}}\label{eqn:upstream}
\end{equation}
The size of the FUV heated region is determined by the column depth required to absorb FUV photons ($N_{\rm FUV}$). In our problem, the disc size is always significantly smaller than where the sonic point in the FUV heated region would occur. As such, the majority of the FUV photons are absorbed in the last scale height at the disc's radius ($R_d$) or:
\begin{equation}
    N_{\rm FUV}\sim \frac{n_{I,d}c^2_{s,FUV}R_d^2}{GM_*} \label{eqn:column}
\end{equation} where $n_{I,d}$ is the density in the FUV heated region at the disc's edge and can be related to the upstream density at the ionization front, assuming the FUV heated region has an approximately hydrostatic density profile\footnote{This is where our argument diverges from \citet{Johnstone1998}, who can't make this approximation due to their large discs.} (Equation~\ref{eqn:hydrostatic}); however, since $R_d \ll R_{IF}$ in this critical case, we only retain the leading order exponential term:
\begin{equation}
    n_{\rm I}\sim n_{I,d} \exp\left(-\frac{GM_*}{c^2_{s,FUV}R_d}\right) \label{eqn:hydrostatic_simple}
\end{equation}
Combining Equations~\ref{eqn:upstream}--\ref{eqn:hydrostatic_simple} we arrive at the condition for the critical flux:
\begin{eqnarray}
    \phi_{\rm crit} &\sim& \alpha_b\frac{N_{FUV}^2 R_g^3}{R_d^4}\exp\left(-\frac{2R_g}{R_d}\right)\left(\frac{c_{s,FUV}}{c_{s,EUV}}\right)^4\nonumber \\ &\sim& 10^{16} ~{\rm photons~s^{-1}~cm^{-2}} ~\exp\left(-\frac{2R_g}{R_d}\right)\label{eqn:euv_crit}
\end{eqnarray}
 We have evaluated this for a solar-mass star where $R_g\sim 100$~AU, with a disc 10~AU in size, $N_{\rm FUV}\sim 10^{21}~{\rm cm}^{-2}$ and an FUV heated sound-speed that is about three times lower than the EUV heated regions. This critical flux is entirely dominated by the exponential term, which for our chosen disc size evaluates to $\phi_{\rm crit}~\sim 3\times 10^7 ~{\rm photons~s^{-1}~cm^{-2}}$. Thus, photoevaporation will be EUV controlled for the expected disc sizes in the Galactic Centre. 
}
\section{Functional fits to the photoevaporation rates}
As stated in the main text, direct use of our new photoevaporation rates would involve computationally intensive multidimensional interpolation of tables. As such, we choose to fit physically motivated functional forms to our photoevaporative rates. We find these rates are generally reasonable to a few percent and, at worst, are wrong by 15\% in corners of parameter space. Our numerical calculated mass-loss rates span a grid in fluxes in the range $10^{14}-10^{17}$ photons s$^{-1}$ cm$^{-2}$, stellar masses in the range 0.5-1.5~M$_\odot$, disc radii in the range 0.5-40~AU and dust-to-gas mass ratios of $10^{-5}-10$. 

We begin by noting that EUV-heated regions are taken to be in ionization-recombination balance. Therefore, the density at the ionization front scales as the square root of the EUV flux \citep{Bertoldi1990,Johnstone1998}. Additionally, in the sub-sonic region, the density profile is hydrostatic (Equation~\ref{eqn:hydrostatic}). Thus, the density in this region should scale exponentially with stellar mass \citep{Adams2004}. Thus, all other things being fixed, the mass-loss should scale with stellar mass and EUV flux as:
\begin{equation}
    \dot{M}_w \propto \sqrt{\phi_{EUV}}\exp\left(-M_*\right)
\end{equation}
We now consider the impact of the disc's size. This has two components: larger discs subtend a larger area to absorb EUV flux over and launch a photoevaporative flow; this should scale in a power-law fashion with disc radii; however, again, in the sub-sonic region, the exponential density profile will exponentially change the density in the outflow. Thus, we choose a functional form:
\begin{equation}
    \dot{M}_w \propto R_d^{d_0}\exp\left[-\left(\frac{10^{d_1}}{R_d}\right)^{d_2}\right]
\end{equation}
to encapsulate the direct impact of the disc's radius. Ignoring the FUV heated outflow and dust's effect, we would roughly expect $d_0=1$, $d_1\sim \log_{10}\left(R_g\right)$ and $d_2=1$. Finally, to include the dust-to-gas ratio, we note again it should be a combination of a power law (due to the change in the size of the FUV heated region) and an exponential function (due to absorption of EUV photons from the $\exp(-\tau)$ term in the ionization rate). Hence we include the dust-to-gas ratio dependence via:
\begin{equation}
    \dot{M}_w \propto X_d^{-a_c}\exp\left[-\left(b_cX_d\right)^{c_c}\right]
\end{equation}
As the point at which you switch from being dominated by the power law or exponential term in the above expression depends on the disc's radius, the coefficients with a sub-script $c$ are functions of the disc's radius. Inspection of their dependencies indicates that they can fit with smoothed, broken power laws of the form:
\begin{equation}
a_c = a_0 + a_1\left(\frac{R_d}{10^{a_3}}\right)^{a_2}    
\end{equation}
and,
\begin{equation}
    x_c = \frac{x_4}{\left[\left(\frac{R_d}{10^{x_3}}\right)^{x_0}+\left(\frac{R_d}{10^{x_3}}\right)^{x_1}\right]^{x_2}}
\end{equation}
where $x$ represents the set of coefficients $\{b,c\}$. 
Normalising our mass-loss rates to a photon flux of $10^{16}$~photons~s$^{-1}$~cm$^{-2}$ and stellar mass of 1~M$_\odot$ our expression for the mass-loss rate becomes:
\begin{eqnarray}
    \dot{M}_w &=& 10^{d_3} \sqrt{\frac{\phi}{10^{16}~{\rm photons~ s}^{-1}~{\rm cm^{-2}}}}\exp\left[-\left(\frac{M_*}{1~{\rm M}_\odot}-1\right)\right]\nonumber \\ &\times& R_d^{d_0}\exp\left[-\left(\frac{10^{d_1}}{R_d}\right)^{d_2}\right]\nonumber \\
    &\times& X_d^{-a_c}\exp\left[-\left(b_cX_d\right)^{c_c}\right]~{\rm g~s}^{-1}\label{eqn:fit}
\end{eqnarray}
where the disc's radius is in cm. The different coefficients are numerically determined by fitting Equation~\ref{eqn:fit} to our numerical determined mass-loss rates. In doing this we find: $a_0 = 0.34153474$, $a_1= 0.55650943$, $a_2=1.3883336$, $a_3=14.99752896$; $b_0 = 0.67174481$, $b_1 = -0.70214445$, $b_2= -1.91599381$, $b_3 = 14.10699905$, $b_4=0.345636$; $c_0 = -0.19865532$, $c_1 = 1.67305016$, $c_2 = 0.50126654$, $c_3=13.96864609$, $c_4 = 1.265315$; $d_0=0.70941993$, $d_1=13.62415834$, $d_2=1.41880752$, $d_3=8.00915806$. These coefficients indicate that the functional forms all transition in shape around a disc radius of $R_g$, as would be expected for any outflow that transitions from sub-critically to super-critically launched. 

\bsp	
\label{lastpage}
\end{document}